\documentclass[draft]{agujournal2019}
\usepackage{apacite}
\usepackage{url}  
\usepackage{amsmath}
\usepackage{graphicx}
\usepackage[percent]{overpic}
\usepackage{lineno}

\draftfalse

\journalname{Space Physics}

\begin{document}

\title{Deriving CME density from remote sensing data and comparison to in-situ measurements}

\authors{M. Temmer\affil{1}, L. Holzknecht\affil{1}, M. Dumbovi\'c\affil{2}, B.  Vr\v{s}nak\affil{2}, N. Sachdeva\affil{3}, S.G. Heinemann\affil{1}, K. Dissauer\affil{1,4}, C. Scolini\affil{5,6}, E. Asvestari\affil{1,7}, A. M. Veronig\affil{1,8}, S. J.  Hofmeister\affil{1}}

\affiliation{1}{Institute of Physics, University of Graz, Austria}
\affiliation{2}{Hvar Observatory, Faculty of Geodesy, University of Zagreb, Croatia}
\affiliation{3}{Climate and Space Sciences and Engineering Department, University of Michigan, USA}
\affiliation{4}{NorthWest Research Associates, Boulder, USA}
\affiliation{5}{Centre for mathematical Plasma Astrophysics (CmPA), KU Leuven, Belgium}
\affiliation{6}{Solar-Terrestrial Centre for Excellence -- SIDC, Royal Observatory of Belgium, Belgium}
\affiliation{7}{Department of Physics, University of Helsinki, Finland}
\affiliation{8}{Kanzelh\"ohe Observatory for Solar and Environmental Research, University of Graz, Austria}

\correspondingauthor{Manuela Temmer}{manuela.temmer@uni-graz.at}

\begin{keypoints}
\item (Interplanetary) Coronal mass ejections
\item CME density evolution
\item Modeling and observations
\end{keypoints}

\begin{abstract}
We determine the 3D geometry and deprojected mass of 29 well-observed coronal mass ejections (CMEs) and their interplanetary counterparts (ICMEs) using combined STEREO-SOHO white-light data. From the geometry parameters we calculate the volume of the CME for the magnetic ejecta (flux-rope type geometry) and sheath structure (shell-like geometry resembling the (I)CME frontal rim). Working under the assumption that the CME mass is roughly equally distributed within a specific volume, we expand the CME self-similarly and calculate the CME density for distances close to the Sun (15--30 Rs) and at 1~AU. Specific trends are derived comparing calculated and in-situ measured proton densities at 1~AU, though large uncertainties are revealed due to the unknown mass and geometry evolution: i) a moderate correlation for the magnetic structure having a mass that stays rather constant ($cc\approx0.56-0.59$), and ii) a weak correlation for the sheath density ($cc\approx0.26$) by assuming the sheath region is an extra mass - as expected for a mass pile-up process - that is in its amount comparable to the initial CME deprojected mass. High correlations are derived between in-situ measured sheath density and the solar wind density ($cc\approx-0.73$) and solar wind speed ($cc\approx0.56$) as measured 24 hours ahead of the arrival of the disturbance. This gives additional confirmation that the sheath-plasma indeed stems from piled-up solar wind material. While the CME interplanetary propagation speed is not related to the sheath density, the size of the CME may play some role in how much material could be piled up. 
\end{abstract}

\section{Introduction}
 
The propagation behavior of coronal mass ejections (CMEs) is governed by different forces at different heliocentric distances. Close to the Sun the propelling Lorentz force dominates, which is reinforced by magnetic reconnection processes \cite<see>[]{chen89,chen96,kliem06,vrsnak08c,vrsnak16}. With time, as the CME moves farther away from the Sun and the magnetic reconnection weakens, CME propagation is predominantly governed by its interaction with the ambient solar wind flow \cite<see e.g.,>[]{cargill96,vrsnak01,temmer11,vrsnak13,sachdeva15}. The magnetohydrodynamic CME drag can be expressed analogously to the aerodynamic drag. It is dependent on the CME geometry, i.e., cross-section and width, density and speed relative to the ambient solar wind \cite<see e.g.,>[]{vrsnak10}. The ram-pressure defined by the CME impact speed and density, is found to be well correlated to the amplitude of sudden storm commencements caused by the rapid compression of the Earth's magnetic field \cite{gonzalez89}. Methods for deriving the CME speed and arrival time at Earth from the CME initial speed close to the Sun have been largely tested \cite<e.g.,>[]{vrsnak13,mays15,sachdeva15,dumbovic18,riley18}. However, the derivation of the CME density using remote sensing data and its evolution to 1~AU has not been investigated yet. 

In-situ measurements of the interplanetary counterpart of CMEs (so called ICMEs; we note that we follow the notation by \citeA{rouillard11} and use the term ICME such to include the shock signature, the sheath and the magnetic structure of the CME) at 1~AU reveal that the density of the sheath region is larger compared to the density within the magnetic ejecta region \cite<e.g., see>[]{masias16}. This can be well explained by the expansion of the low plasma beta magnetic structure that dilutes the plasma material ejected from the Sun. As the CME propagates and expands, solar wind material piles-up in front of the driver forming the sheath region as was reproduced by MHD simulations \cite<e.g.,>[]{siscoe08}. ICME sheath regions reveal in general different characteristics compared to the magnetic ejecta regions. In particular they show low alpha to proton and Fe/O ratios, characteristic of the ambient solar wind, and stronger turbulence \cite<e.g., see>[]{kilpua17b,owens18}. Evidence that the sheath becomes more prominent in interplanetary space is also given by a relative increase of the sheath durations from Mercury to Earth \cite{janvier19}. Recent studies using remote sensing data suggest that in the early CME evolution there is no significant mass pile-up ahead of the CME \cite{howard18}. Observed mass increases up to 20 solar radii (R$_{\rm s}$) is most probably due to outflows from the solar surface \cite{bemporad10,bein13,temmer17,veronig19}. 

A changing CME mass has consequences for the drag force acting on the CME in interplanetary space \cite{vrsnak08a,cargill04}, and with that modifies the CME propagation time, impact speed, and energy input to the magnetosphere \cite{takahashi17}. Moreover, a better understanding and quantification of CME mass density close to the Sun is important to provide accurate inputs for numerical models in order to properly simulate CME propagation and to predict Space Weather. Usually the CME mass density is taken as a constant default value based on observational studies of streamer densities. For example ENLIL \cite{odstrcil99} uses the so-called CME cloud density (dcld) parameter that by default is four times larger than typical mean values in the ambient fast wind \cite<see>[]{mays15}, and for EUHFORIA \cite{pomoell18} a value of 10$^{-18}$ kg/m$^3$ is taken corresponding to dcld=2, i.e., half of the density used by default in ENLIL (dcld=4).

With the launch of the Solar TErrestrial RElations Observatory \cite<STEREO;>[]{kaiser08} and its SECCHI instrument suite \cite{2008howard_SECCHI} in 2006, the interplanetary evolution of the CME structure in white-light can be measured seamlessly over the Sun--Earth distance range and is used to derive the CME kinematics all the way from Sun to 1~AU. Tracking density and mass from Sun to Earth is a quite complex undertaking. In a case study, \citeA{savani13} showed a qualitative comparison of an estimated mass profile for the sheath structure using white-light heliospheric image data that resembled quite well the in-situ density measurements. In the current study, we aim for a better understanding of the density evolution of the ICME sheath and magnetic structure. Furthermore, we derive the CME density close to the Sun in order to feed CME propagation models. For a set of 29 well-observed multi-viewpoint STEREO CME-ICME pairs, we perform a statistical study covering the derivation of the CME kinematics, deprojected mass, and 3D geometry parameters from which we calculate the CME volume and density, and compare the results to in-situ measurements at 1~AU.   

The paper is structured as follows. In Section~\ref{sec:methods} the data set and the methodology are presented. Section~\ref{sec:res} gives the analysis and results. In Sections~\ref{sec:discuss} and~\ref{sec:sum} the results are discussed and summarized. In the Appendix we provide for all studied events physical parameters that can be used as input for CME propagation models.

\section{Data and Methods}\label{sec:methods} 
The study is based on a sample of 29 well-observed CME-ICME pairs that occurred in the period December 2008--August 2014 (extracted parameters for each event are given in the Appendix in Table~\ref{tab:SHINE-list}). The events were selected so as to have reliable stereoscopic observations plus a spacecraft encounter from which we derive clear in-situ plasma and magnetic field measurements. In that respect, we restricted our selection to CMEs that were observed by both STEREO satellites (Ahead and Behind) and revealed in-situ signatures at L1 by Wind. To correctly link the CME-ICME pairs we first estimated the CME arrival time at 1~AU using the drag-based model by \citeA{vrsnak13} (available as ESA tool under: http://swe.ssa.esa.int/web/guest/graz-dbm-federated) with the CME initial speed, distance, and angular extent in the ecliptic plane as obtained from the GCS reconstruction \cite<see also>[]{dumbovic19}. Applying a time window of $\pm$18 hours, centered at the calculated arrival time, we searched in the in-situ data for the corresponding ICME characteristics. All the results are cross-checked with ready lists from \citeA<>[to which we refer further on as R\&C list]{richardson10}, \citeA{wood17}, and HELCATS (Heliospheric Cataloguing, Analysis and Techniques Service) WP4 catalogue (LINKCAT).

For the linked CME-ICME events verified in this way we obtain from stereoscopic observations their 3D parameters close to the Sun such as propagation direction, speed, geometry, and furthermore the deprojected mass \cite<see>[]{colaninno09} and in-situ characteristics for near-Earth space. The geometry and volume of a CME together with its deprojected mass is further used to estimate the particle density of an ICME at 1~AU applying to the CME volume a self-similar expansion with different expansion rates. For comparing calculated and measured in-situ densities as well as other parameters, we use the bootstrap method. We apply 10$^4$ repetitions, i.e., the bootstrapping procedure re-samples the original data 10.000 times from which 10.000 statistical samples are created in order to derive the Spearman median correlation coefficient and an 80\% empirical confidence interval \cite{efron1979_bootstrap,efron93_bootstrap}.

\subsection{Parameter definition close to the Sun}\label{sec:2.1} 
\begin{figure}
\noindent\includegraphics[width=\textwidth]{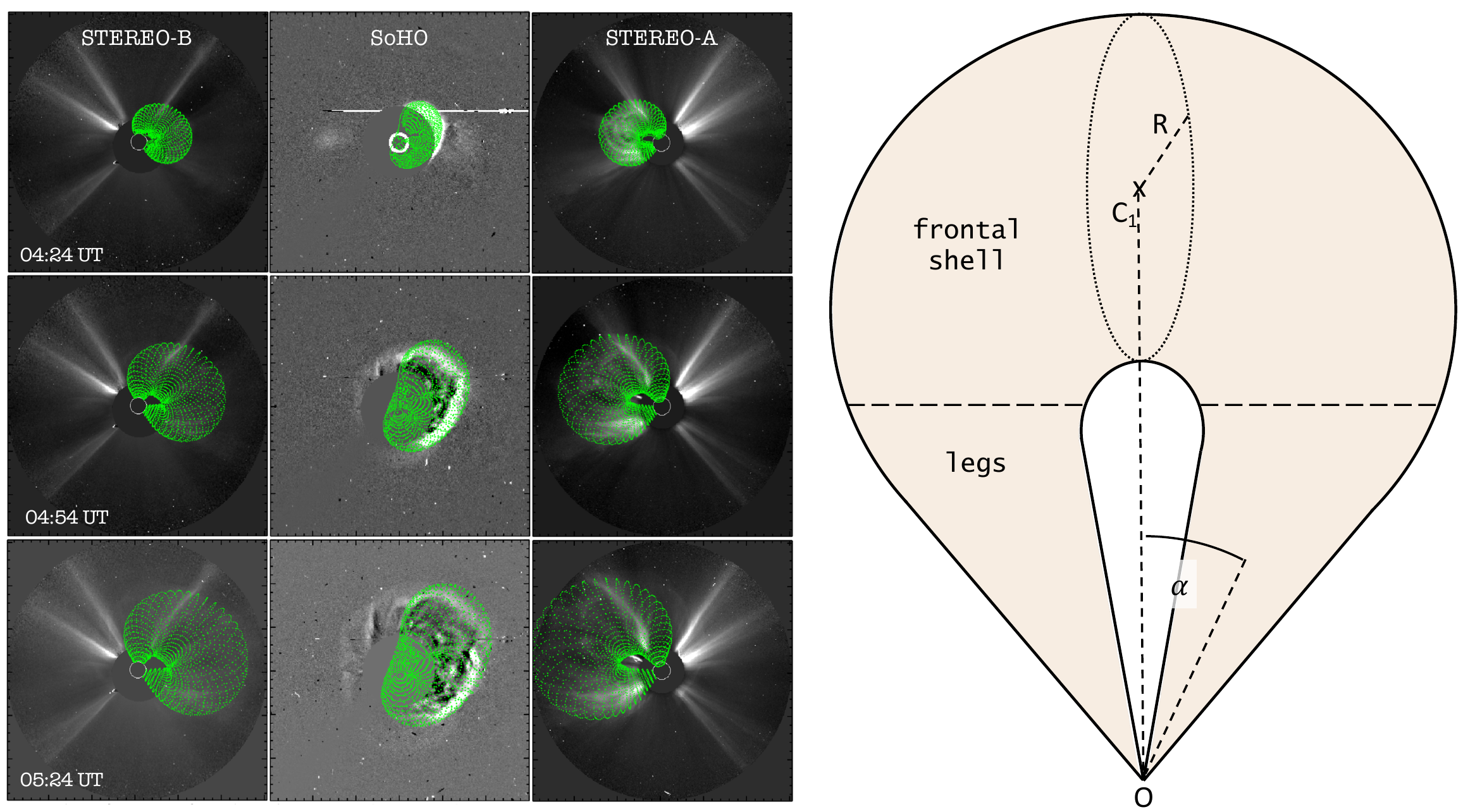}
\caption{Left: Example of the 3D geometry for the CME from August 4, 2011 reconstructed using the GCS model by \citeA{thernisien11} for three vantage points (COR2 from STEREO-A and -B, and LASCO/C3). Right: Sketch of the GCS model giving the parameters required for calculating the CME volume ($\alpha$ is the angular half-width, $C_1$ is the apex center, $R$ is the cross-section radius at the apex). GCS ``full volume'', as given in the text, refers to the frontal shell plus legs.}
\label{fig:gcs}
\end{figure}

The CME 3D geometry at a distance range of about 15~R$_{\rm s}$ is reconstructed with the graduated cylindrical shell model \cite<GCS; see>[]{2009thernisien,thernisien11} using combined white-light coronagraph data from 2 or 3 different viewpoints of SECCHI/COR2 aboard STEREO-A and -B and LASCO/C2/C3 aboard the Solar and Heliospheric Observatory \cite<SOHO;>[]{1995soho}. The GCS model represents an idealized geometry of a CME flux rope as a hollow croissant-shaped mesh that is manually fitted to white-light data with the presumed magnetic structure of the CME. The left panels in Figure~\ref{fig:gcs} show a GCS reconstructed flux rope for the CME event on August 4, 2011 covering the height range 8 to 16~R$_{\rm s}$ at three time steps. As this reconstruction is subjective, we aim to avoid bias and use for the study GCS reconstructions performed independently by three different groups: a) N. Sachdeva thesis \cite{sachdeva19}, 12 events using three s/c viewpoints; b) UNIGRAZ, 10 events using three s/c viewpoints; c) HELCATS WP3 COR2 Catalogue KINCAT (which builds on the work undertaken during the EU FP7 AFFECTS project), 7 events using two s/c viewpoints from STEREO-A and -B. 

The CME volume ($V$) is calculated as function of height/distance using the GCS parameters $\alpha$ (angular half-width) and $\kappa$ (aspect ratio) as described in \citeA{holzknecht18}. The right panel of Figure~\ref{fig:gcs} sketches the derivation of those parameters with $\alpha$, $\kappa=(\overline{OC_1})/R$ (where $O$ is the center of the Sun, $C_1$ is the apex center, and $R$ is the cross-section radius at apex), and $h$ is the height of the leading edge given as $h=(\overline{OC_1})+R$ according to \citeA{thernisien11}. The entire GCS volume consists of the frontal shell and two legs. Over the LASCO field of view up to $h=30$~R$_{\rm s}$, a self-similar expansion is assumed. From this we define the full volume of the expanding GCS flux rope, $V_{\rm fr}$, close to the Sun.

The CME deprojected mass in the corona, $m_{\rm dp}$, is derived at $\sim$15~R$_{\rm s}$ by using combined white-light coronagraph data from SECCHI/COR2 aboard STEREO-A and -B. The method assumes that the difference in the white-light excess brightness, as measured by the two STEREO spacecraft, is due to the plane of sky projection of an optically thin source. The brightness excess is converted into electron excess, hence mass, assuming a composition of 90\% hydrogen and 10\% helium. The masses calculated from the two vantage points are then corrected for the CME propagation direction. This is done by varying consistently the masses derived from each vantage point until they yield the same result. The technique and application is described in \citeA{colaninno09} and \citeA{bein13}. The region of interest from which the mass is derived, was defined either manually by drawing the flux rope boundary or, as given for the HELCATS catalogue, by using the GCS model boundaries \cite{savani13,pluta19}. A cross-check between both approaches revealed no systematic differences in the derived mass.

\subsection{Parameter definition for near-Earth space}
In-situ plasma and interplanetary magnetic field measurements (5-minute averaged) are taken from the Wind spacecraft and its Solar Wind Experiment \cite{ogilvie95} and Magnetic Field Experiment \cite{lepping95}. We extract from the R\&C list the times of the shock arrival and start/end times of the magnetic ejecta (ME) structure \cite<based primarily on plasma and magnetic field observations; for more details see>[]{richardson10} from which we define the sheath and ME region. In this study we primarily use the arithmetic mean and standard deviation of in-situ measured plasma density and speed, over the duration of each structure (see Appendix A for additional parameters that were extracted).

\begin{figure}\centering
\noindent\includegraphics[width=0.7\textwidth]{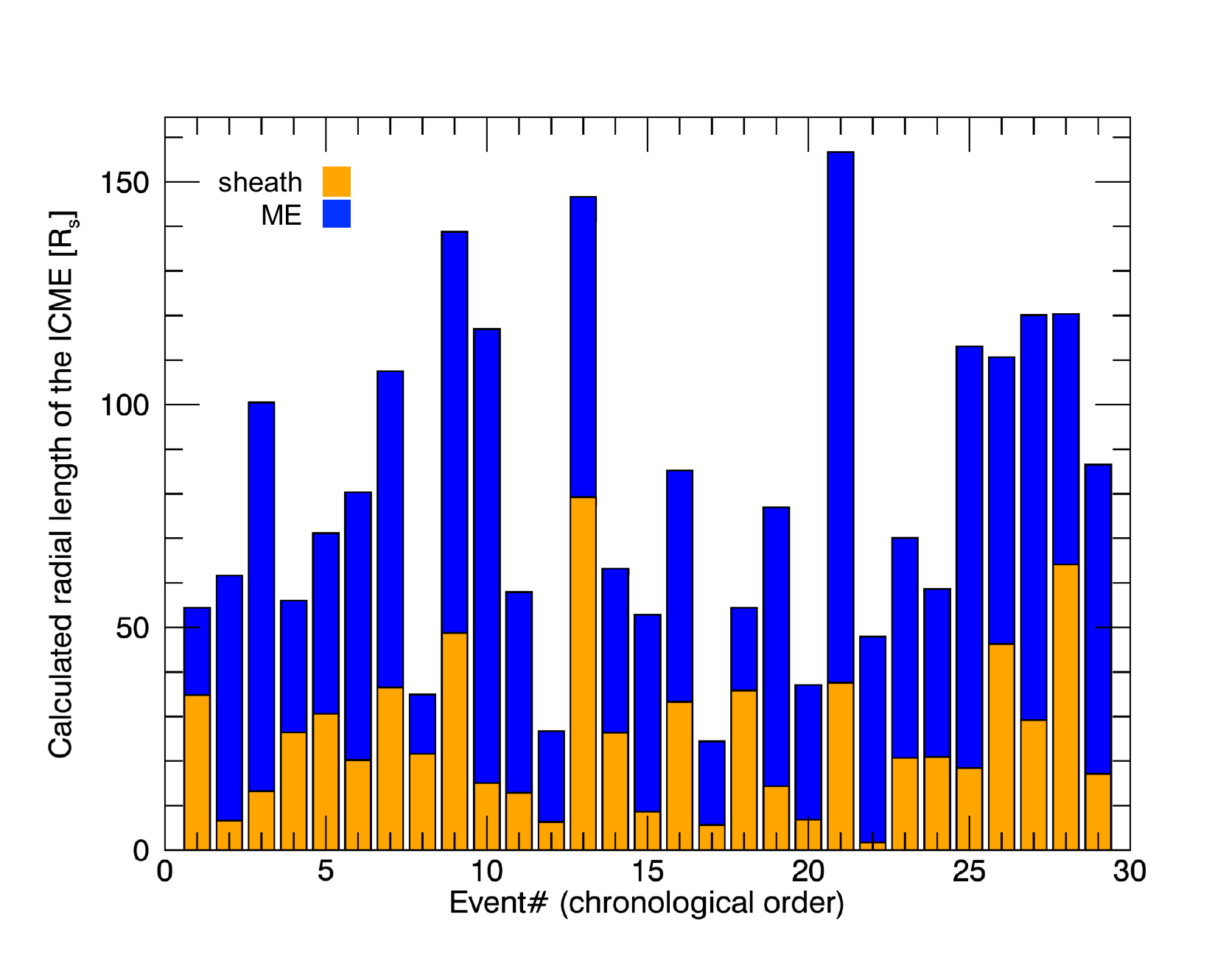}
\caption{Radial size of ICME structures (sheath and magnetic ejecta - ME) at 1~AU 
for each event that is used for the volume calculation. As reference point the shock arrival time is used which is set to $h=215$~R$_{\rm s}$. Sheath is given in orange, ME in blue.}
\label{fig:length}
\end{figure}

\begin{figure}
\noindent\includegraphics[width=\textwidth]{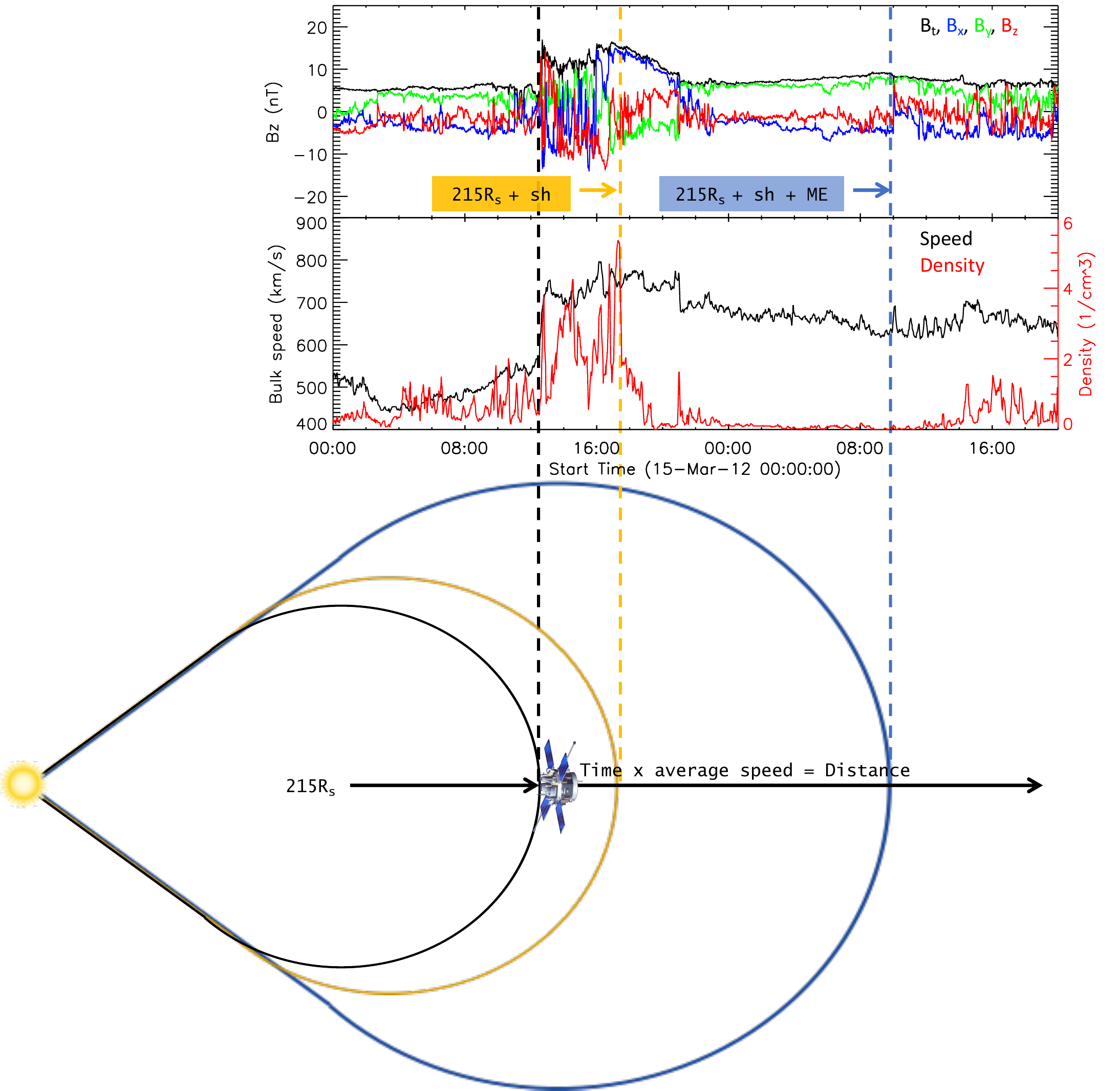}
\caption{Sketch of volume calculation in relation to the in-situ measured ICME structures (sheath, ME). Different radial sizes of the GCS geometries are used for the volume calculation. Black: $V_{\rm 215}$ derived from a radial size of 215~R$_{\rm s}$; Orange: $V_{\rm sh}$ derived from the radial size 215~R$_{\rm s}$+sheath; Blue: $V_{\rm ME}$ is derived from the radial size 215~R$_{\rm s}$+sheath+ME.}
\label{fig:fig2}
\end{figure}

As the CME expands in interplanetary space it propagates over the spacecraft with the sheath region being detected first, followed by the ME. Figure~\ref{fig:length} gives the radial size of the sheath and ME structure for each of the events. They are calculated by multiplying the duration of the sheath (shock arrival time until start time of ME) and duration of the ME (start time of ME until end time of ME), respectively, by the average speed of each structure (we note that this method assumes constant expansion over the averaging time interval). The sheath is of shorter duration/size compared to the ME, hence relates also to a smaller volume. For calculating densities of the ME and sheath structure we use therefore different volumes. Figure~\ref{fig:fig2} sketches the CME volume derivation using the in-situ measurements of an ICME at 1~AU. The arrival time of the ICME shock is used as reference point and sets $h$ to 215~R$_{\rm s}$ (average 1~AU distance). The derived distances are used for calculating the full CME volume (frontal shell plus legs; cf., right panel of Figure~\ref{fig:gcs}) for the sheath and ME structure. We define 1) $V_{\rm 215}$ as the volume up to the in-situ measured shock by expanding the GCS flux rope to $h=215$~R$_{\rm s}$, 2) $V_{\rm sh}$ as the sheath volume by expanding the GCS flux rope to $h=215$~R$_{\rm s}$ + sheath distance, and 3) $V_{\rm ME}$ as the ME volume by expanding the GCS flux rope to $h=215$~R$_{\rm s}$ + sheath distance + ME distance. 

For calculating the densities from the volume and mass, we assume that 1) the observed initial deprojected mass from white-light data represents mostly plasma within the ME volume, $m_{\rm ME} \approx m_{\rm dp}$; 2) the mass is uniformly distributed within the volume structures; 3) the ME structure may have some mass exchange with the interplanetary medium, i.e., the ME experiences either mass loss due to erosion or mass growth due to reconnection with the interplanetary magnetic field \cite<see e.g.,>[]{dasso07,manchester14,2015ruffenach}; 4) the sheath region acquires additional mass, due to mass accumulation at the CME front while propagating in interplanetary space. \citeA{deforest13} found from an observational case study a CME mass increase by 60\%, that would be consistent with swept up solar wind material in the sheath region. According to these assumptions we vary the mass of the ME structure and the sheath region. The mass of the ME structure, $m_{\rm ME}$, is varied by $\pm$25\% of the initial deprojected mass $m_{\rm dp}$. The sheath region mass, $m_{\rm sh}$, is calculated for 0.5, 1 and 1.5 times $m_{\rm dp}$. The entire CME mass at 1~AU yields $m_{\rm total}=m_{\rm ME}+m_{\rm sh}$. For the volume we consider self-similar expansion with different expansion trends by varying the expansion factor $x$ when calculating the flux rope radius with $R(h)=R_0(h/h_0)^x$, where $R_0$ and $h_0$ refer to the starting value at $h$=15~R$_{\rm s}$. Different expansion rates are tested by varying $x$ in the range of 0.8--1.1 (in steps of 0.1) according to results from observational studies \cite<e.g.>[]{1998bothmer+schwenn,2007leitner,2008demoulin_solphys,2012gulisano,vrsnak19}. For comparison with in-situ data we derive the proton number density, $n_{\rm p}$, given in cm$^{-3}$ by assuming abundances of 90\% hydrogen and 10\% helium.

\section{Results}\label{sec:res}

\subsection{CME density profile for 15--30~R$_{\rm s}$}
\begin{figure}
\noindent\includegraphics[width=\textwidth]{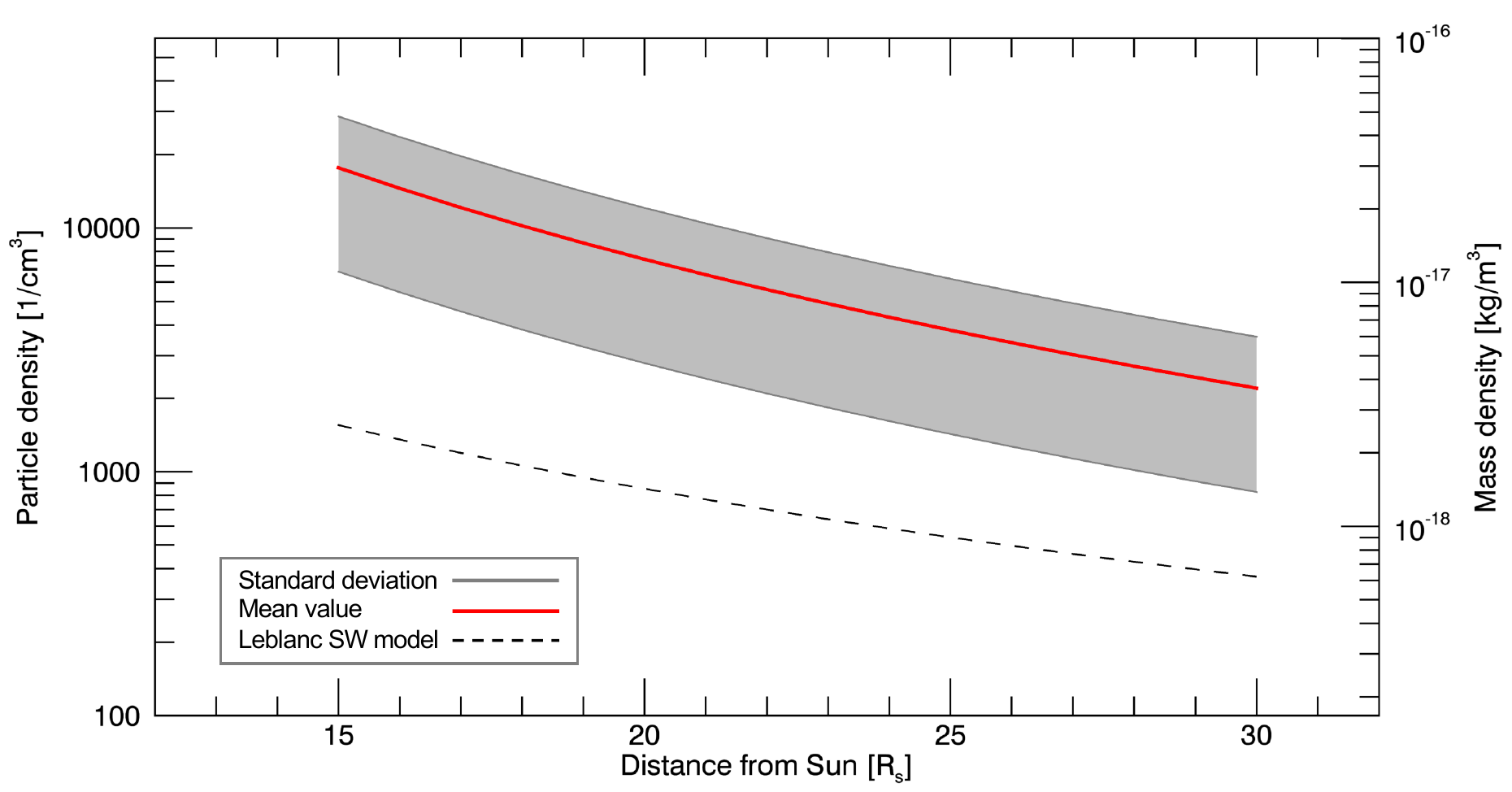}
\caption{CME density over the distance range 15--30~R$_{\rm s}$. We assume that the CME flux rope (full volume) expands in a self-similar manner ($x$=1.0) and derive for all 29 events the mean distance-density profile (red line) together with its standard deviations (black lines), as calculated from $\alpha$, $\kappa$ and $m_{\rm ME}=m_{\rm dp}$ values for the individual events. As comparison we give the density profile from the empirical solar wind model by \citeA{leblanc98} with the density at 1~AU normalized to $n_p=7~$cm$^{-3}$ (dashed line). }
\label{fig:dens_sun}
\end{figure}

Close to the Sun, for the distance range up to 30~R$_{\rm s}$ from observational studies no mass pile-up is reported \cite{howard18}. Assuming that the estimated deprojected mass represents mostly plasma within the ME volume, Figure~\ref{fig:dens_sun} shows for the entire sample of 29 events the average CME density profile, $\rho(h)$, and standard deviation over the distance range $h=15-30$~R$_{\rm s}$ derived by $\rho(h)=m_{\rm dp}/V_{\rm fr}(h)$ and assuming self similar expansion ($x$=1) and a uniform plasma distribution within the CME structure. The plot presents the CME proton number density, $n_{\rm p}$, in particles per cm$^{3}$ (left y-axis) and the corresponding mass density in kg~m$^{-3}$ (right y-axis). On average, over the distance range 15--30~R$_{\rm s}$ the CME particle density lies in the range of $\sim(2.2-17.6)\times10^{3}$ cm$^{-3}$ (mass densities with $\sim(3.6-29)\times10^{-18}$ kg~m$^{-3}$). For comparison, we give the solar wind density profile applying the relation by \citeA{leblanc98}. From this we derive a ratio between CME and solar wind density that decreases from $\sim$11 at $h=15$~R$_{\rm s}$ to $\sim$6 at $h=30$~R$_{\rm s}$. 

\subsection{Comparing calculated and in-situ measured densities using the full volume}
\begin{figure}
\noindent\includegraphics[width=\textwidth]{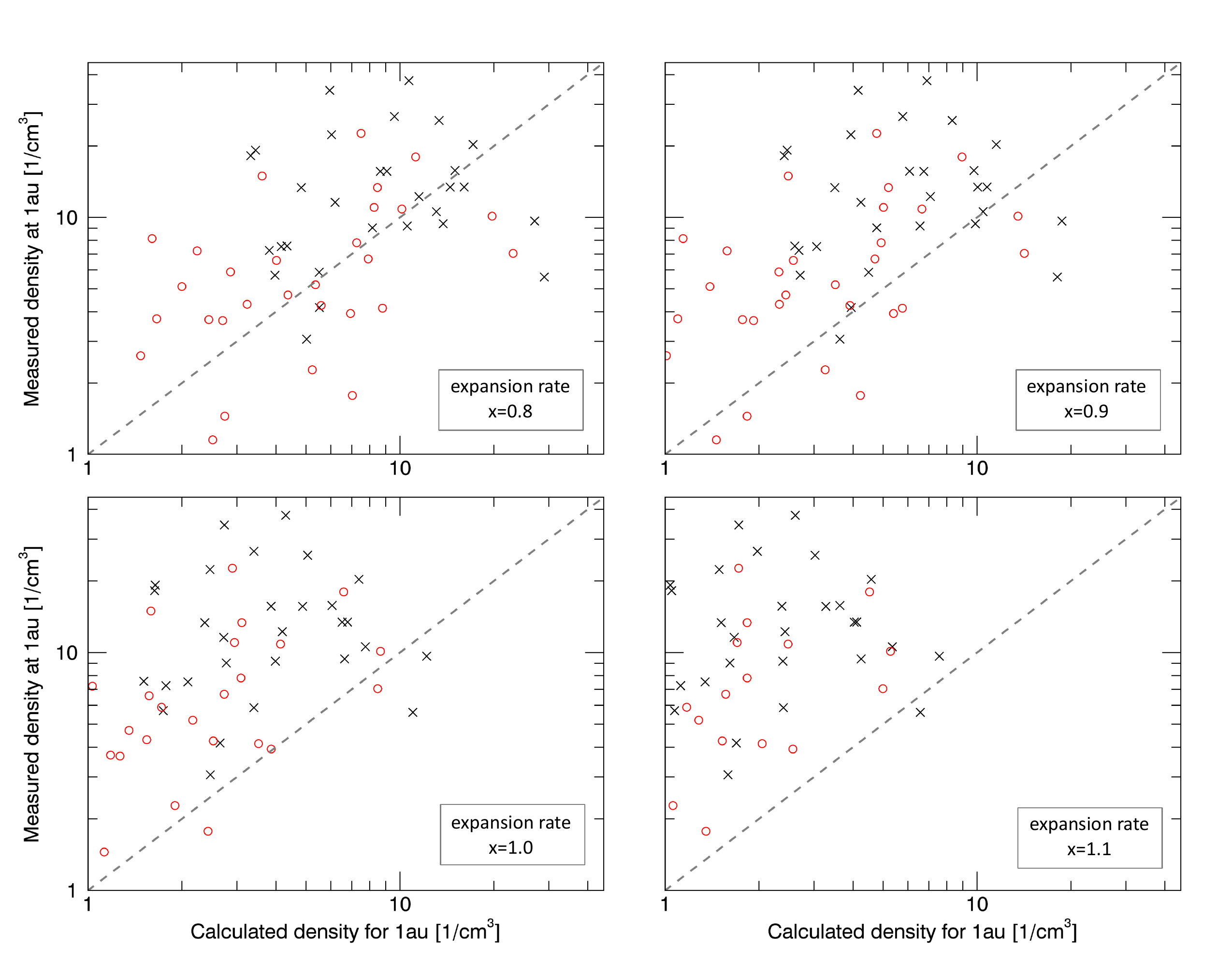}
\caption{Log-log plot of the calculated versus measured particle density using the full volume for each structure and different expansion factors (see legend). ME is marked by red circles, sheath by black crosses. Gray dashed line gives the $x=y$ equality. The used input mass is $m_{\rm sh} = m_{\rm ME} = m_{\rm dp}$. }
\label{fig:vol215}
\end{figure}

Figure~\ref{fig:vol215} gives the comparison between calculated (x-axis) and measured (y-axis) proton plasma densities at 1~AU, separately for the sheath and ME structure. For the volume derivation of the sheath ($V_{\rm sh}$) and ME ($V_{\rm ME}$) we vary the expansion factor in the range $x$=0.8--1.1. The results are derived in a first approximation by simply applying a mass of $m_{\rm sh}$ = $m_{\rm ME} = m_{\rm dp}$. For low expansion factors, $x$=0.8--0.9, the calculated plasma densities are obtained to be of the same order of magnitude as the in-situ measured ones. These findings first of all support that the CME geometry/volume derivation and mass calculation based on remote sensing image data are physically meaningful. The sheath region does not consist of coronal CME plasma, but likely piled-up preceding solar wind plasma. Therefore, the sheath region has to be considered as an extra mass separate to the coronal CME mass \cite<c.f.>[]{kilpua17b}. For larger volumes using $x$=1.0--1.1, the calculated densities tend to get underestimated.

\begin{figure}
\noindent\includegraphics[width=1.1\textwidth]{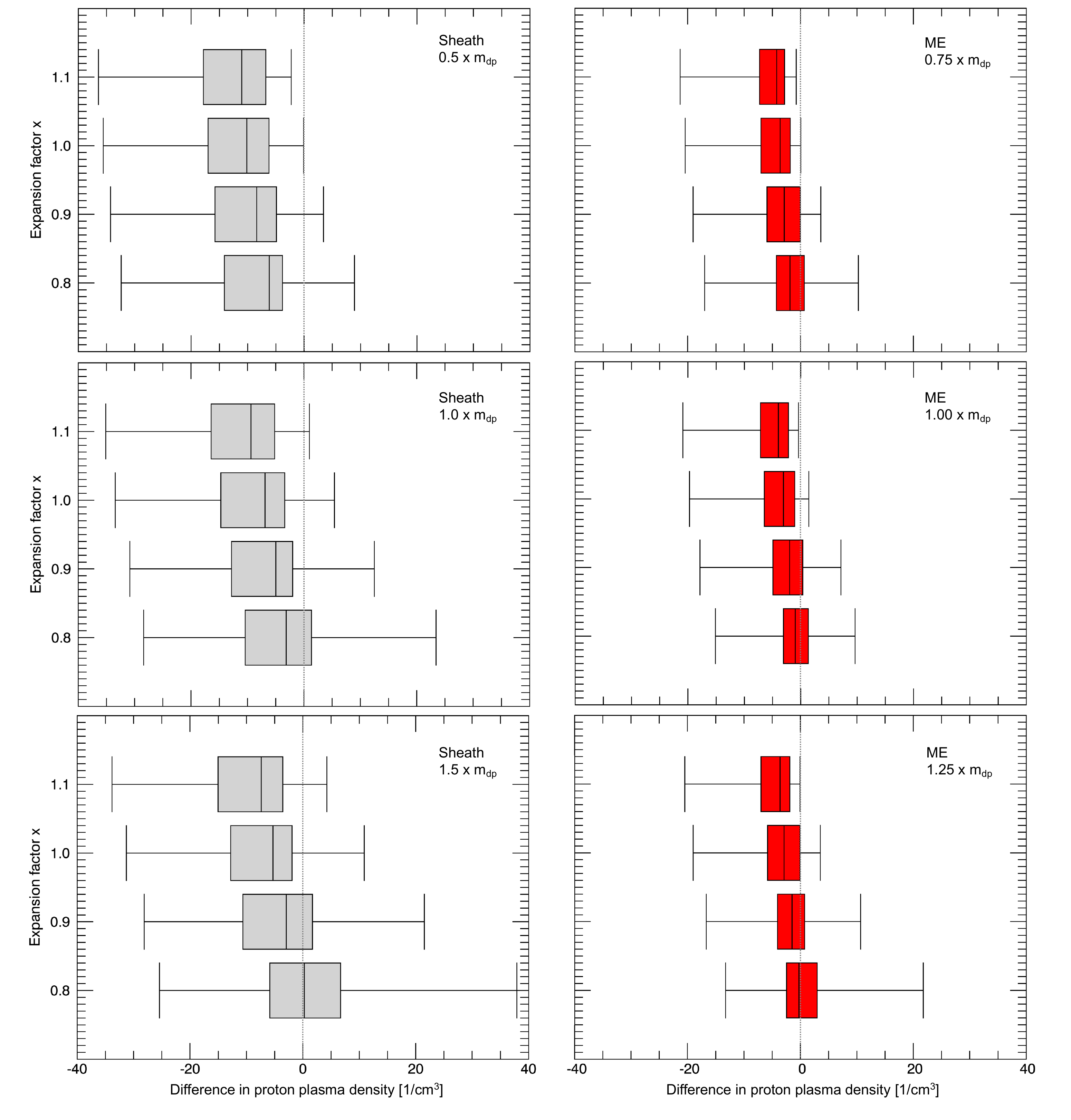}
\caption{Boxplots showing the differences between calculated and observed proton plasma density (left panel sheath; right panel ME region) using different expansion factors. Different multiples of the derived deprojected mass (see plot legends) are used to calculate from the full volume the proton plasma density separately for the sheath and ME region. The whiskers give the minimum and maximum values, and the box itself is confined by the lower and upper quartile, with the vertical line within marking the median.}
\label{fig:vol215_box}
\end{figure}

In a next step we vary the mass of the different structures for taking into account evolutionary processes, just like erosion or mass accumulation in interplanetary space. Figure~\ref{fig:vol215_box} shows the derived differences between the calculated and observed plasma densities using different expansion factors and input mass, represented as box and whiskers plot containing the minimum, lower quartile, median, upper quartile, and maximum for each value array. Table~\ref{tab:med-ci} (two left columns) gives the median values and standard deviation for the derived differences between the calculated and measured plasma density values. In general, the calculated plasma densities tend to be underestimated when compared with in-situ measurements. This effect increases for larger volume expansions ($x>0.9$). For the sheath region, using 50\% of the initial deprojected mass the results are not satisfying. However, when increasing the mass to $1.5\times m_{\rm dp}$ and using $x$=0.8, the results improve showing differences between calculated and measured densities that are distributed around zero. For the ME region, mass inputs of $0.75-1.25\times m_{\rm dp}$ result in a fair match with the observations with best results for a constant mass or increased mass and low volume expansion ($x$=0.8--0.9). Table~\ref{tab:cc} (two left columns) summarizes the resulting Spearman median correlation coefficients and 80\% confidence intervals from the bootstrapping method separately for sheath and ME structure applying different expansion factors. We note that different mass inputs do not affect the rank-order correlation coefficient. We find that the correlations between measured and calculated particle densities for each structure at 1~AU differ strongly. While the ME region shows a weak correlation ($cc\approx0.47-0.49$) though with large spread, for the sheath region almost no correlation is found ($cc\approx0.11-0.19$).

\subsection{Comparing calculated and in-situ measured densities using a reduced volume}
\begin{figure}
\noindent\includegraphics[width=\textwidth]{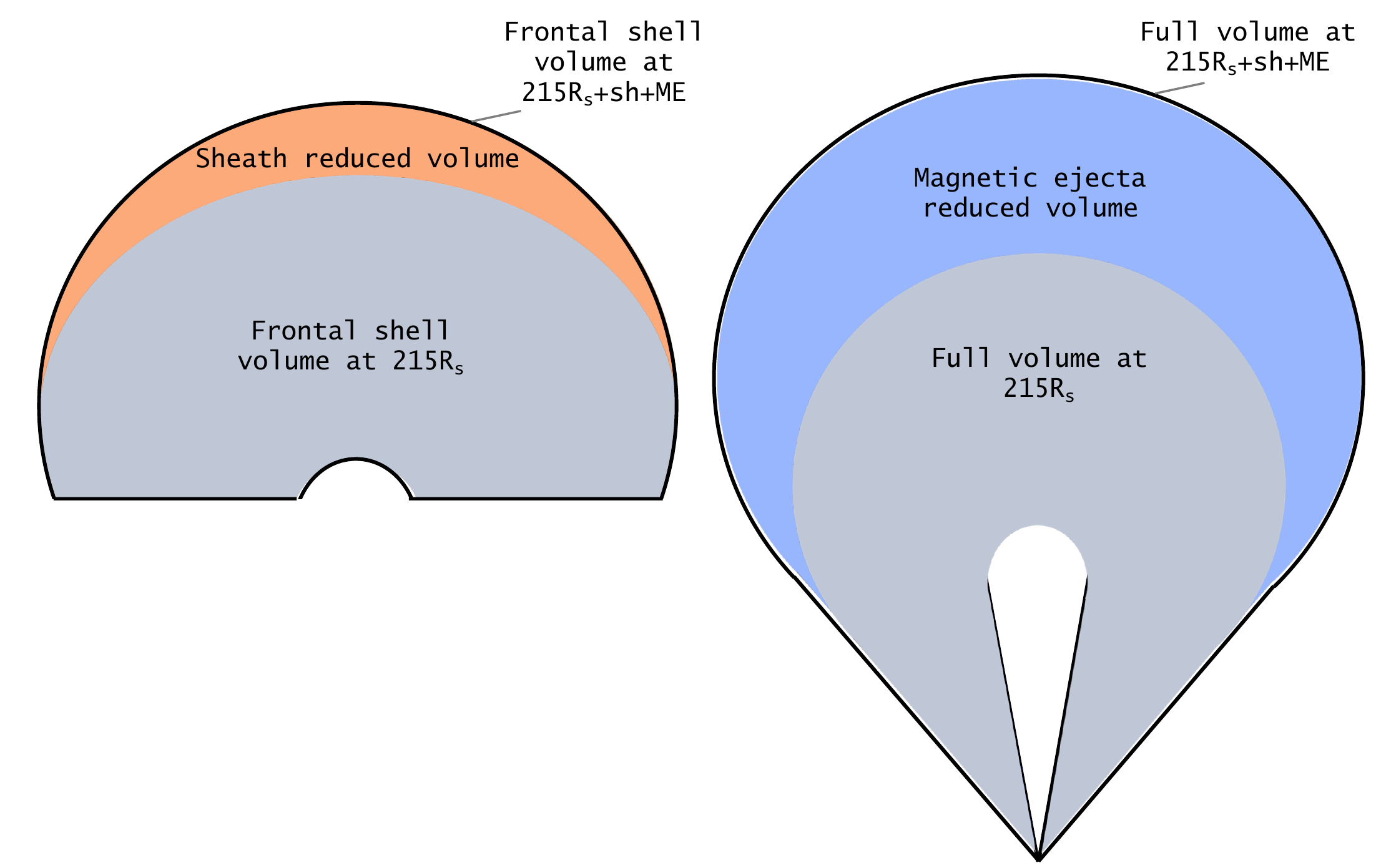}
\caption{Sketch of reduced volume derivation for the sheath region (left) and the ME region (right) that is derived by subtracting the CME volume up to 215R$_{\rm s}$. }
\label{fig:cartoon}
\end{figure}

The ME seems rather well reproduced with the simple GCS flux rope geometry. We also find that lower expansion factors tend to give better results in terms of more symmetrically distributed values around the line of equality between calculated and measured densities. So far, in our simple approach, we have assumed that the plasma of the sheath region is distributed within a flux-rope type geometry, same as the ME but of smaller volume. However, the sheath region presumably has not a flux-rope type but more likely a shell-like geometry. For improving the statistics we test various geometries for the volume derivations (e.g., subtracting differently expanding volumes from each other) and find best results for a ``reduced volume" which is sketched in Figure~\ref{fig:cartoon}. First, the sheath volume is based solely on the frontal shell geometry of the GCS reconstructed flux rope (see also Figure~\ref{fig:gcs}), while for the ME the basis is still the entire GCS flux rope. Then from both structures corresponding volumes for $h=215$~R$_{\rm s}$ are subtracted (cf., Figure~\ref{fig:cartoon}), meaning that we calculate the reduced volumes by $V_{\rm ME,red} = V_{\rm ME}-V_{\rm 215}$ and $V_{\rm sh,red}=V_{\rm ME,frontal}-V_{\rm 215,frontal}$.

\begin{table}[]
\caption{Median values and standard deviation for the derived differences between measured and calculated density values (see Figures~\ref{fig:vol215_box} and \ref{fig:vol_reduced_out_box}). Different expansion rates $x$ and volumes (full or reduced) were used for the sheath and magnetic ejecta (ME) region. The mass indices, $m_x$, refer to a: $m_{\rm sh} = 0.5 \times m_{\rm dp}$ and $m_{\rm ME} = 0.75 \times m_{\rm dp}$;  b: $m_{\rm sh} = 1.0 \times m_{\rm dp}$ and $m_{\rm ME} = 1.0 \times m_{\rm dp}$; c: $m_{\rm sh} = 1.5 \times m_{\rm dp}$ and $m_{\rm ME} = 1.25 \times m_{\rm dp}$; }
\label{tab:med-ci}
\centering
\begin{tabular}{cc|cccc}
\hline
$x$  & $m_x$  & \multicolumn{2}{c}{full volume} & \multicolumn{2}{c}{reduced volume} \\
    &   & \multicolumn{1}{c}{sheath} & \multicolumn{1}{c}{ME}  & \multicolumn{1}{c}{sheath}           & \multicolumn{1}{c}{ME}              \\  
    \hline
    & a & $-6.2\pm9.3$    & $-1.9\pm5.2$  & $-5.3\pm15.6$     & $+1.1\pm11.1$    \\
0.8 & b & $-3.1\pm10.9$    & $-1.0\pm5.8$  & $-1.2\pm26.8$     & $+2.0\pm14.8$    \\
    & c & $+0.3\pm13.2$    & $-0.3\pm6.6$  & $+5.0\pm38.7$     & $+4.4\pm18.6$    \\
\hline
    & a & $-8.4\pm9.0$    & $-2.9\pm4.8$  & $-8.3\pm11.3$     & $-1.1\pm6.9$    \\
0.9 & b & $-5.0\pm9.8$    & $-2.0\pm4.9$  & $-5.1\pm16.4$     & $-0.3\pm8.9$    \\
    & c & $-2.9\pm11.0$    & $-1.5\pm5.2$  & $-2.9\pm22.4$     & $+1.0\pm11.1$    \\
\hline
    & a & $-10.1\pm8.8$   & $-3.6\pm4.7$  & $-10.6\pm9.6$    & $-2.9\pm5.0$    \\
1.0 & b & $-6.9\pm9.2$    & $-3.1\pm4.7$  & $-8.1\pm11.7$     & $-1.8\pm5.8$    \\
    & c & $-5.3\pm9.8$   & $-2.9\pm4.7$  & $-5.9\pm14.3$     & $-0.9\pm6.8$    \\
\hline
    & a & $-11.0\pm8.8$   & $-4.2\pm4.8$  & $-11.4\pm9.0$    & $-3.7\pm4.5$    \\
1.1 & b & $-9.4\pm8.9$    & $-4.0\pm4.7$  & $-10.5\pm9.8$    & $-3.3\pm4.6$    \\
    & c & $-7.4\pm9.2$    & $-3.6\pm4.6$  & $-9.2\pm10.8$     & $-3.1\pm4.9$   
\end{tabular}
\end{table}

\begin{figure}
\noindent\includegraphics[width=1.1\textwidth]{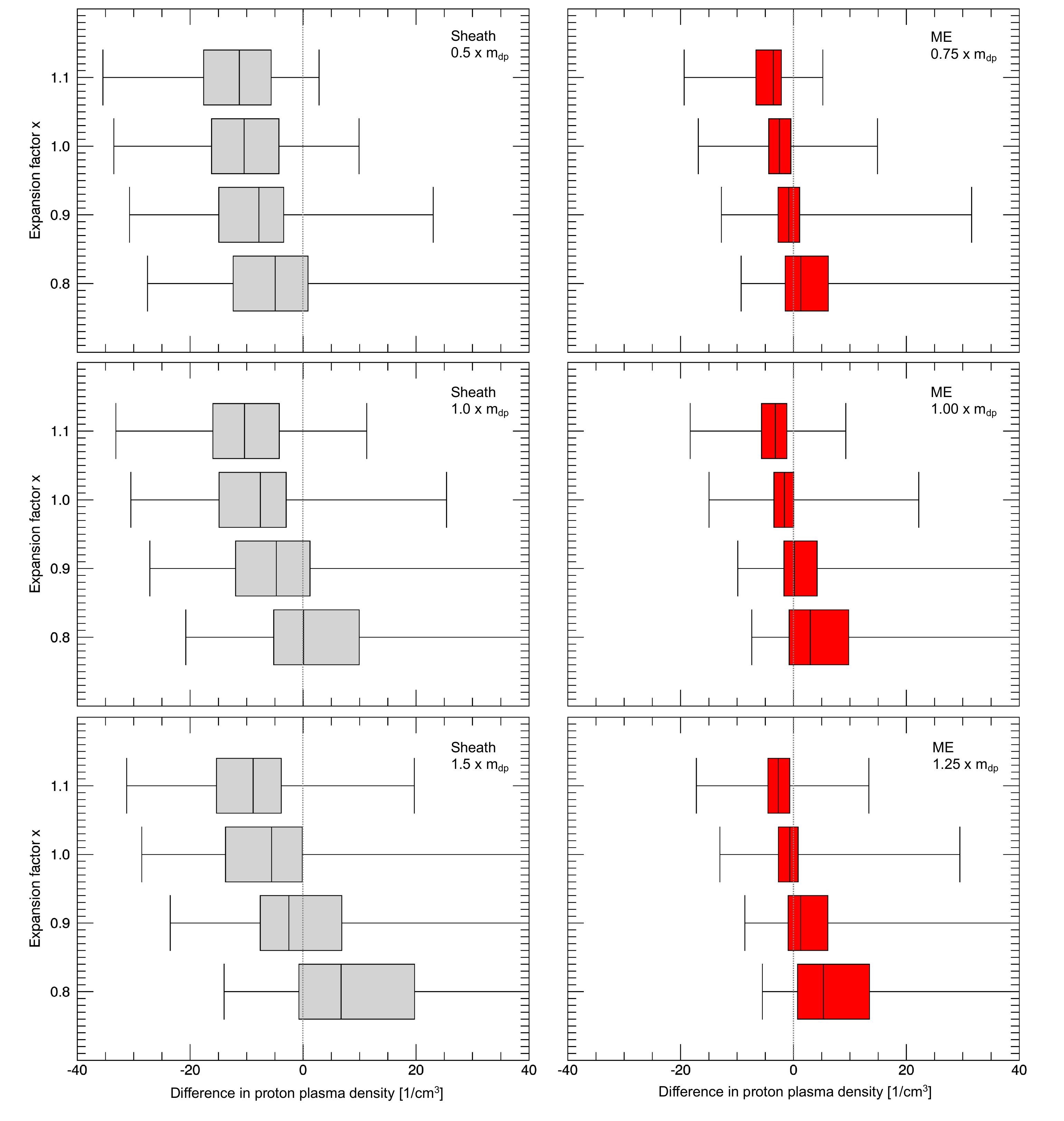}
\caption{Same as Figure~\ref{fig:vol215_box} but using the reduced volume for the calculation of sheath and ME plasma density.}
\label{fig:vol_reduced_out_box}
\end{figure}

In general, by applying the reduced volumes, some events yield calculated densities that largely exceed the measurements (maximum is $\sim$38 particles cm$^{-3}$). These are three events where we used $x$=0.8 (in-situ shock arrival times: 11-Apr-10 13:04UT, 04-Feb-11 01:55UT, 17-Jun-11 02:41UT) and two events with $x$=0.9 (in-situ shock arrival times: 04-Feb-11 01:55UT and 17-Jun-11 02:41UT). 
Figure~\ref{fig:vol_reduced_out_box} shows the differences between calculated and observed plasma densities, represented as box and whiskers plots (containing the minimum, lower quartile, median, upper quartile, and maximum for each value array). Table~\ref{tab:med-ci} (two right columns) gives the median and standard deviation for the derived differences between calculated and measured plasma density values. In comparison to the results for the full volume (Figure~\ref{fig:vol215_box}), the derived median values for sheath and ME are closer to zero, however, the spread in the values increases. For the ME structure the lower and upper quartile get narrower, while for the sheath region they increase. Best results for the sheath region are obtained for an extra mass which is comparable to the initial deprojected mass and for a CME geometry of weak expansion ($x$=0.8). For the ME region, equally good results are obtained for either a decreased or constant mass and low expansion ($0.75-1.0\times m_{\rm dp}$; $x$=0.9) or increased mass ($1.25\times m_{\rm dp}$) and a volume that is expanding with $x$=1.0. The two right columns in Table~\ref{tab:cc} give for the reduced volume the Spearman median correlation coefficients between calculated and measured densities and 80\% confidence intervals from the bootstrapping method separately for sheath and ME structure for the different expansion factors. For the sheath region, by applying a reduced volume, we obtain a slightly improved correlation with $cc\approx0.26$ and for the ME structure we get $cc\approx0.56-0.59$.

\begin{table}
\caption{Spearman median correlation coefficients ($cc$) and 80\% confidence interval (CI 80\%) resulting from the bootstrapping analysis (see Figures~\ref{fig:vol215_box} and \ref{fig:vol_reduced_out_box}). Different expansion rates $x$, and volumes (full or reduced) were used for the sheath and magnetic ejecta (ME) region. For more details see text.}
\label{tab:cc}
\centering
\begin{tabular}{c|cccc}
\hline
$x$ & \multicolumn{2}{c}{full volume} & \multicolumn{2}{c}{reduced volume} \\
 & sheath  & ME & sheath & ME  \\
 \hline
0.8 & 0.19  & 0.47 & 0.25 & 0.56  \\	
(CI 80\%) & (0.42;$-$0.07) & (0.64;0.26) & (0.47;0.01) & (0.71;0.36)  \\
0.9	& 0.14 & 0.47 & 0.25 & 0.57 \\
(CI 80\%) & (0.38;$-$0.11) & (0.63;0.26) & (0.48;0.01) & (0.71;0.38) \\
1.0 & 0.14 & 0.48 & 0.26 & 0.58  \\
(CI 80\%) & (0.37;$-$0.11) & (0.65;0.28) & (0.48;0.01) & (0.72;0.39) \\
1.1 & 0.11 & 0.49 & 0.26 & 0.59  \\
(CI 80\%) & (0.35;$-$0.13) & (0.65;0.28) & (0.48;0.01) & (0.73;0.41) \\
\hline
\end{tabular}
\end{table}

\begin{figure}
\noindent\includegraphics[width=1.1\textwidth]{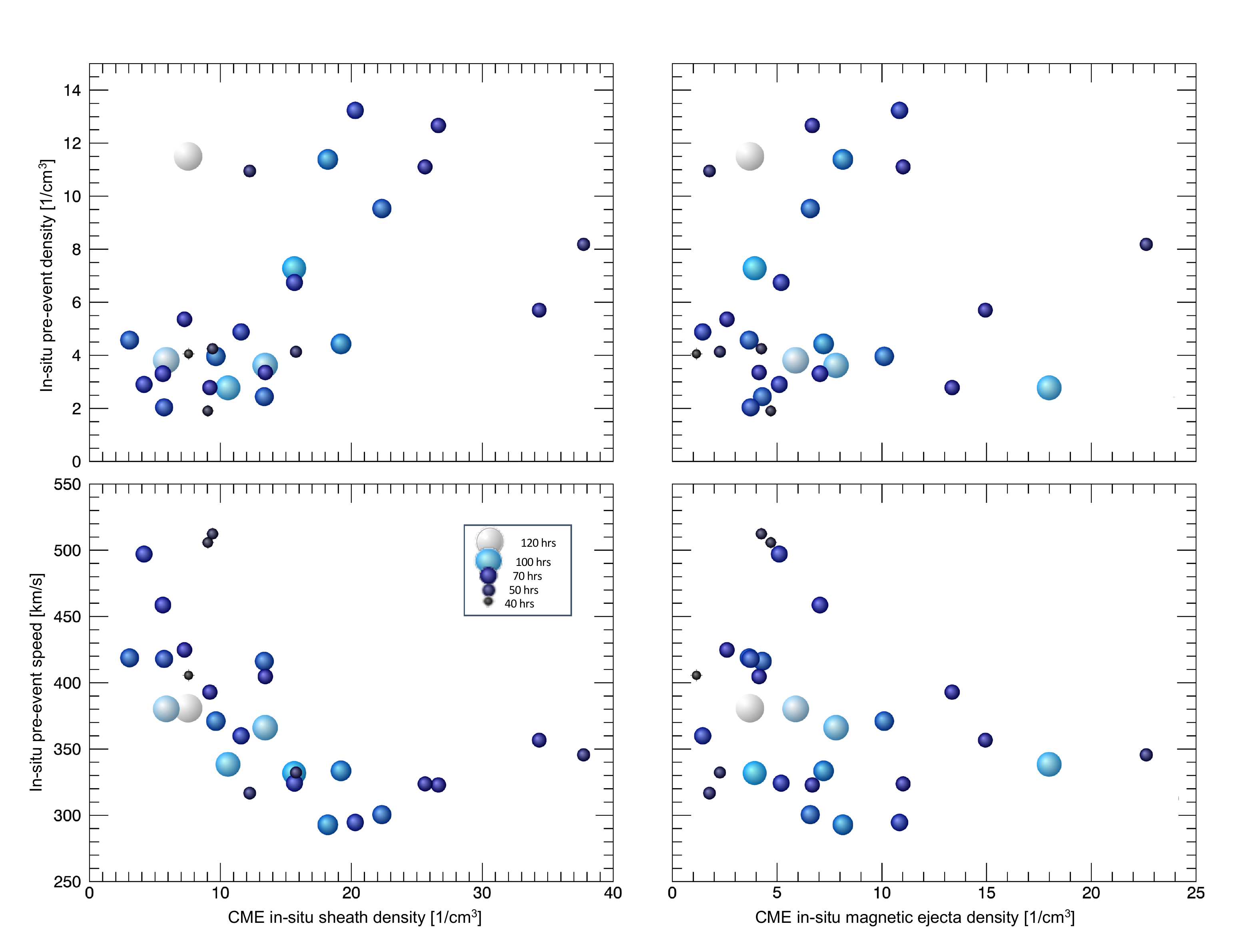}
\caption{Top: In-situ measured solar wind density 24 hours before shock arrival versus average density of the sheath (left) and magnetic ejecta (right). Bottom: In-situ measured solar wind speed 24 hours before shock arrival versus in-situ measured average density of the sheath (left) and magnetic ejecta (right). Color coded bubbles mark the propagation duration (transit time) of the CME in hours (see legend).}
\label{fig:sh_mc_pre}
\end{figure}

In order to obtain more conclusive results about the interplay between ambient solar wind, CME characteristics and mass/density evolution, we further investigate the relation between observational parameters derived from remote sensing and in-situ data. Figure~\ref{fig:sh_mc_pre} shows the relation between measured density in the sheath and ME region versus solar wind plasma characteristics measured 24 hours before the arrival of the disturbance (pre-event density and pre-event speed). In addition, we color-code each parameter pair with the CME transit time (time difference between CME LASCO/C2 first appearance and ICME shock arrival at 1~AU, given in hours). In contrast to the ME region, the sheath structure shows a much stronger dependence on the conditions ahead of the disturbance. The highest anti-correlation is found between sheath density and pre-event speed with $cc=-0.73$, and the largest positive correlation between sheath-density and pre-event density with $cc=0.56$ (ME density and pre-event speed: $cc=-0.30$; ME density and pre-event density: $cc=0.09$). Table~\ref{tab:pre-dens-cc} summarizes the Spearman median correlation coefficients and 80\% confidence intervals from the bootstrapping method between the ICME sheath and ME density against solar wind plasma parameters (density, speed) and the total magnetic field, measured 24 and 48 hours before the arrival of the disturbance. The dependencies clearly decrease when relating the parameters to the conditions 48 hours ahead of the disturbance. We find no dependencies on the CME transit time, hence, interplanetary CME speed, and no correlation to the pre-event magnetic field.

We further inspect how the size of the CME is related to the amount of piled-up solar wind material and other CME characteristics that are important for investigating the drag force in interplanetary space. Table~\ref{tab:alpha-kappa} gives the statistical results from the bootstrapping analysis between the CME geometry parameters $\alpha$ (angular half-width) and $\kappa$ (aspect ratio) as derived from GCS (as defined in Section~\ref{sec:2.1}) versus CME deprojected mass, transit time, and in-situ measured particle densities for the sheath and ME region. We find a moderate anti-correlation between geometry and transit time ($cc=-0.60$ for $\alpha$ and $cc=-0.62$ for $\kappa$), while for the deprojected mass a moderate correlation with $\alpha$ ($cc=0.67$) and a weak correlation with $\kappa$ is obtained ($cc=0.39$). Comparing the geometry parameters with in-situ measurements we get a weak to moderate correlation between $\alpha$ and $\kappa$ and the sheath density ($cc=0.19$ and $0.30$) and a very weak one for the ME density ($cc=-0.13$ and $0.16$). Opposite to the ME, the sheath formation seems to be related to the CME size.

\begin{table}
\caption{Spearman median correlation coefficient derived from the bootstrapping analysis for the pre-event density measured 24 and 48 hours (pre24 and pre48), respectively, before shock arrival versus sheath and ME density (de), speed (v) and total magnetic field (B) taken from in-situ measurements.}
\label{tab:pre-dens-cc}
\centering
\begin{tabular}{c|cccccc} 
\hline
in-situ & pre24-de & pre48-de & pre24-v & pre48-v & pre24-B & pre48-B\\
\hline
sheath-de &  0.56 & 0.50 & $-$0.73 & $-$0.66 & 0.10 & $-$0.06\\
(CI 80\%)  & (0.70;0.37) & (0.67;0.30) & ($-$0.63;$-$0.80) & ($-$0.52;$-$0.77) & (0.33;$-$0.16) & (0.18;$-$0.30)\\
ME-de &  0.08 & 0.09 & $-$0.30 & $-$0.19 & $-$0.18 & $-$0.20 \\
(CI 80\%)  &  (0.30;$-$0.16) & (0.30;$-$0.13) & ($-$0.09;$-$0.48) & (0.02;$-$0.40) & (0.06;$-$0.42) & (0.036;$-$0.43)\\
\hline
\end{tabular}
\end{table}

\begin{table}
\caption{Spearman median correlation coefficient derived from the bootstrapping analysis between GCS geometry parameters $\alpha$ (angular half-width), $\kappa$ (aspect ratio) as defined in Section~\ref{sec:2.1}, the GCS source region longitude (lon) and latitude (lat)  and CME parameters from remote sensing and in-situ data against measured sheath density (sheath de) and ME density (ME de).}
\label{tab:alpha-kappa}
\centering
\begin{tabular}{c|cccc}
\hline
GCS parameter & transit time & $m_{\rm dp}$ & sheath de & ME de  \\
\hline
$\alpha$  &  $-$0.60            & 0.67      & 0.19              & 0.16  \\
(CI 80\%) & ($-$0.42;$-$0.74)  & (0.77;0.53) & ($-$0.06;0.43) & ($-$0.10;0.40) \\
$\kappa$  &  $-$0.62            & 0.39      & 0.30          & $-$0.13  \\
(CI 80\%) &  ($-$0.42;$-$0.76) & (0.58;0.16)& (0.08;0.51)      & ($-$0.38;0.10) \\
lon & $-$0.41 & 0.22 & 0.30 & 0.08 \\
(CI 80\%) &  ($-$0.59;$-$0.17) & ($-$0.03;0.45) & (0.05;0.51) & ($-$0.18;0.32)\\
lat &  $-$0.08 & 0.08 & 0.11 & $-$0.01 \\
(CI 80\%) & ($-$0.35;$-$0.22) & ($-$0.18;0.32) & ($-$0.14;0.36) & ($-$0.26;0.24) \\
\end{tabular}
\end{table}

\section{Discussion}\label{sec:discuss}
The CME mass/density is a parameter that directly connects to the CME propagation behavior in interplanetary space. CMEs of high density will be less affected by the drag from the ambient solar medium compared to less dense ones. With that, the mass evolution of a CME in interplanetary space influences strongly the propagation duration and speed. Using for a sample of 29 CME-ICME pairs observational data from combined remote sensing and in-situ measurements together with geometry modeling efforts, we assess the CME mass/density evolution from Sun to Earth. 

The geometry and volume of a CME together with its mass is derived from stereoscopic remote sensing data covering a field of view up to $\sim$15~R$_{\rm s}$. This is further used to estimate the particle density of an ICME at 1~AU applying a self-similar CME volume expansion with different expansion rates ($x$=0.8--1.1). We treat the sheath region as extra mass (as first approach we simply use the same amount as the initial mass) and keep the ME initial mass and sheath mass constant as the CME expands. With that we find for $x$=0.8--0.9 a rather good agreement between the ICME calculated and in-situ measured ME density. This supports that the available techniques are reasonable and adequate for deriving mass and volume by using remote sensing image data.

We show a novel approach for computing the CME density by applying GCS reconstruction for CMEs. By calculating the CME volume and expansion in interplanetary space, together with the deprojected mass, the density is derived for various distances from the Sun. As the obtained parameters are based on simplistic geometric modeling of a flux-rope-type CME, it is not surprising that the differences between calculated and observed densities are large. Nevertheless, the statistical results yield trends that allow to draw some conclusions. We derive a moderate relationship between calculated and in-situ measured densities with $cc=0.47-0.49$ using a full volume and $cc=0.56-0.59$ for a reduced volume. This finding may indicate that the ME structure might be reproduced by a flux rope-like geometry. Statistically, the smallest differences between calculated and observed ME densities are derived for a low expansion factor and rather constant mass. This hints towards that the majority of the mass expelled from the Sun lies within the CME flux rope structure and during CME propagation the mass within that flux rope might be rather constant with variations of about $\pm$25\%. The sheath region clearly behaves differently and the geometry most likely deviates from a flux rope shape. We derive only weak correlations between calculated and in-situ measured sheath densities when using different types of volume estimates and the highest correlation is $cc\approx0.26$ for a reduced rim-like volume structure which is similarly shaped as the CME front. The median differences between calculated and observed sheath densities are larger compared to the magnetic ejecta results and increase when reducing the mass input. This gives indication that the sheath needs to be treated as considerable extra mass.

We find a moderate anti-correlation between the GCS source region longitude and transit time, i.e., the more west the source region the shorter the propagation duration \cite<fast CMEs are deflected eastward, slow ones westward; since we observe a plasma pile-up, our study covers mainly fast CMEs; see also>[]{wang04,sudar16}. A weak positive correlation ($cc\approx0.30$) with the measured density is obtained, hence, the more eastward a CME is launched the higher the in-situ measured density. This might hint towards different levels of compression in dependence of the source region location. However, we note that the CME propagation direction might change in interplanetary space (deflection).

Significant relations are found between in-situ measured sheath density and pre-event solar wind conditions (speed, density). Moreover, the geometry of a CME tends to be better related to the measured sheath density than to the ME density. This gives further evidence that the sheath region is largely composed of ambient solar wind material as the CME propagates through interplanetary space. The sheath density is found to be higher when the CME propagates in slow solar wind (cf.\,top and bottom panel to the left in Figure~\ref{fig:sh_mc_pre}). This could also be interpreted in terms of compression which is larger for CMEs propagating in slow solar wind \cite{owens18}. Slow solar wind is more dense compared to fast streams, hence, CMEs propagating in slow solar wind are of higher sheath densities as there is more material ahead of the CME to be piled up.

Our findings are supported by previous studies such as \citeA{deforest13} who reports a CME mass increase in interplanetary space of the order of 60\% and \citeA{janvier19} who finds for aligned events covering Mercury and Earth measurements, a relative increase of the duration of the sheath compared to the ME structure. Typically, the variation of CME mass as consequence of the interaction with the background solar wind is described by the concept of virtual mass \cite{cargill96,cargill04}. More explicitly, the ``snow plough'' model considers solar wind mass pile-up at the CME front in analogy to a plough \cite{tappin06}.

The average CME density ratio with the background solar wind is $\sim$6 at 15~R$_{\rm s}$ and $\sim$11 at 30~R$_{\rm s}$  \cite<see also>[]{ontiveros09}. For 21.5~R$_{\rm s}$ (inner boundary for heliospheric models) we obtain $\sim$7$\pm$4. Keeping in mind that the background solar wind density is based on slow solar wind, our results suggest dcld values higher than the default values currently used for CME propagation models. It is known that variations of the density parameter strongly influence the CME propagation time and impact speed \cite{mays15,werner19}. Values derived from observations should be used as cross-check and to restrict the choice of free parameters in the propagation models. In the Appendix A, model input CME parameters for the sample of 29 events are given for the distance range 21.5~R$_{\rm s}$.

\section{Conclusions}\label{sec:sum}
The current study is based on observational data from remote sensing and in-situ instruments combined with GCS fitting. The simple geometry assumptions used and the unknown mass and geometry evolution in interplanetary space, makes it unfeasible to exactly pin down the complex relations between CME expansion, mass evolution and interaction with the solar wind. However, from the statistical analysis we derive specific trends from which we conclude as follows. 

CME magnetic structure: \\
\begin{itemize}
   \item our results are in agreement with a scenario in which the major part of CME mass close to the Sun, as measured from remote sensing white-light data, lies within a closed magnetic structure (presumably the flux rope)
    \item during propagation some mass exchange with the ambient solar wind could be possible 
    \item measured in-situ densities within the magnetic ejecta could be explained by a CME volume expanding self-similarly with $x\approx0.9-1.0$
\end{itemize}

CME sheath region: \\
\begin{itemize}
    \item the sheath region forms and consists of piled-up interplanetary solar wind material
    \item the amount of piled-up mass depends on a) the prevailing density and solar wind flow speed in interplanetary space ahead of the CME, and b) the CME size (wider CMEs act as piston leading to a stronger mass pile-up compared to narrow CMEs acting like a bow shock where plasma can more easily flow around). 
\end{itemize}

With the newly launched satellites Parker Solar Probe \cite<PSP;>[]{fox16} and Solar Orbiter \cite{mueller20}, we will have more information on the CME density for various distances. PSP orbits in the near future will access the LASCO/C3 coronagraphic field of view ($<$30~R$_{\rm s}$) and approach the Sun as close as 10~R$_{\rm s}$. With that we will have the opportunity to actually measure CME density characteristics over the distance range 15--30~R$_{\rm s}$ and to compare with the derived results from this study.

\appendix

\section{CME-ICME pairs - event parameters}
Table~\ref{tab:SHINE-list} gives for the 29 CME-ICME pairs under study relevant parameters that may be used as input for modeling. For identification of the CME-ICME pairs we give the CDAW catalogue time based on LASCO observations \cite{yashiro04} and the ICME disturbance arrival time according to the R\&C list \cite{richardson10}. We list from GCS reconstructions (based on remote sensing data from two or three different vantage points) the following parameters: longitude, latitude in Stonyhurst coordinates, the tilt as the angle of the flux rope axis with respect to the equatorial plane, and the geometry parameters $\alpha$ and $\kappa$ from which the volume is calculated \cite{holzknecht18}. The deprojected mass is derived at a distance of about 15~R$_{\rm s}$ using combined STEREO-A and -B COR2 data. For the distance of 21.5~R$_{\rm s}$ (0.1~AU) we give the CME speed and time (kinematics are derived from GCS reconstructions covering several time steps, and making a linear extrapolation from the last two data points to estimate the values for 21.5~R$_{\rm s}$). The density is derived from the deprojected mass divided by the CME volume applying $h=21.5$~R$_{\rm s}$. From in-situ measurements we give the average speed over the sheath region (to estimate the impact speed at 1~AU distance) and the average ME speed, as well as the average sheath and ME particle density. We also list the calculated density based on the reduced CME volume for sheath and ME region applying $x=0.9$. To feed ensemble CME propagation models, we give general estimates of uncertainties empirically derived from the analysis performed during this study: $\pm$30~min in timing, $\pm$50--200~km/s for speed, 10\% for $\alpha$ and $\kappa$, 30\% for the deprojected mass and density, $\pm$10$^{\circ}$ for longitude and latitude, and $\pm$ 20$^{\circ}$ in tilt.

\begin{sidewaystable}
\tiny
\caption{Parameters derived for each of the CME events under study (no.) including a flag (q) marking the GCS fit as very easy (1), easy (2), and hard (3) to perform. We give the CDAW catalogue observation time, GCS reconstruction parameters lon(gitude) [$^{\circ}$], lat(itude) [$^{\circ}$] , tilt [$^{\circ}$] , $\alpha$ (rad), and $\kappa$ (rad), the deprojected mass $m_{\rm dp}$ [g], and information by whom the information was provided (HC - HELCATS; UG - UNIGRAZ; NS - Nishtha Sachdeva). For the distance of 21.5~R$_{\rm s}$ we give the extrapolated CME speed and time, and the calculated mass density [kg/m$^3$]. From in-situ measurements we list the ICME disturbance arrival time according to the R\&C list \cite{richardson10}, the average sheath and ME speed [km/s], the average proton number density for sheath and ME. The last two columns give the calculated proton number density based on the reduced CME volume applying $x$=0.9.}
\label{tab:SHINE-list}
\begin{tabular}{rccccccccccccccccccc}
no.(q) & \multicolumn{8}{c|}{Remote sensing of the Sun and GCS results} & \multicolumn{3}{|c|}{21.5R$_{\rm s}$ (0.1~AU)} &\multicolumn{5}{|c|}{In-situ measurements at 215R$_{\rm s}$ (1~AU)} & \multicolumn{2}{|c}{Calculated}  \\
 & \multicolumn{1}{c}{LASCO} & lon & lat & tilt & $\alpha$ & $\kappa$ & $m_{\rm dp}$ & by &  \multicolumn{1}{c}{time}  & \multicolumn{1}{c}{$v$}    & \multicolumn{1}{c}{$\rho$} & Disturbance  & $\overline{v}_{\rm sh}$ &  $\overline{v}_{\rm ME}$ & $\overline{n_p}_{\rm sh}$ & $\overline{n_p}_{\rm ME}$ & $\overline{n_p}_{\rm sh}$ & $\overline{n_p}_{\rm ME}$   \\

1(3) & 2008/12/12 0525     & 4                     & 5                     & 51                     & 0.23                & 0.27       & 4.50E15 & HC   & 2008/12/12 1500  & 470             & 1.36E-17                               & 2008/12/16 0800 & 350 & 340           & 15.7   & 3.9    & 22.8     & 11.3     \\
2(3) & 2009/12/16 0430     & $-$2                  & 7                     & $-$6                   & 0.39                & 0.31       & 2.20E15 & HC   & 2009/12/16 1230  & 370             & 4.81E-18                               & 2009/12/19 1000 & 430  & 380        & 3.0    & 3.7    & 5.8      & 3.8      \\
3(2) & 2010/04/03 1033     & 3                     & $-$29                 & 2                      & 0.42                & 0.29       & 6.04E15 & UG   & 2010/04/03 1350  & 900             & 1.42E-17                               & 2010/04/05 0826 & 720   & 650          & 9.4    & 4.2    & 9.0      & 5.9      \\
4(2) & 2010/04/08 0454     & $-$2                  & $-$9                  & $-$29                  & 0.57                & 0.19       & 7.52E15 & NS   & 2010/04/08 0900  & 500            & 3.08E-17                               & 2010/04/11 1304 & 430 & 410           & 9.7    & 10.1   & 39.9     & 28.3     \\
5(2) & 2010/05/24 1406     & 13                    & 8                     & $-$10                  & 0.24                & 0.48       & 3.20E15 & HC   & 2010/05/24 0240  & 390             & 4.36E-18                               & 2010/05/28 0258 & 370 & 360            & 19.2   & 7.2    & 4.3      & 2.8      \\
6(1) & 2010/06/16 0635     & $-$17                 & 3                     & $-$33                  & 0.17                & 0.26       & 2.35E15 & NS   & 2010/06/16 2330  & 430           & 7.97E-18                               & 2010/06/20 2000 & 390   & 360        & 5.9    & 5.9    & 9.3      & 3.8      \\
7(3) & 2010/10/26 0200     & 18                    & $-$25                 & $-$55                  & 0.52                & 0.26       & 7.14E15 & NS   & 2010/10/26 1730  & 450            & 1.86E-17                               & 2010/10/30 1015 & 380   & 340          & 13.4   & 7.8    & 10.4     & 7.2      \\
8(3) & 2011/01/30 2008     & $-$40                 & $-$12                 & $-$20                  & 0.20                & 0.26       & 5.60E15 & HC   & 2011/01/30 2150  & 310             & 1.84E-17                               & 2011/02/04 0155 & 370  &410            & 10.6   & 18.0   & 56.3     & 25.8     \\
9(2) & 2011/02/15 0236     & 0                     & $-$11                 & 53                     & 0.37                & 0.46       & 6.83E15 & UG   & 2011/02/15 0630  & 700         & 8.95E-18                               & 2011/02/18 0130 & 540 & 470           & 11.6   & 1.4    & 3.4      & 2.4      \\
10(1) & 2011/03/03 0548     & 8                     & $-$23                 & 8                      & 0.38                & 0.35       & 3.13E15 & NS   & 2011/03/03 1300  & 510            & 5.83E-18                               & 2011/03/06 0331 & 530 & 440           & 4.2    & 5.1    & 3.0      & 2.0      \\
11(3) & 2011/06/02 0745     & 42                    & 12                    & 55                     & 0.42                & 0.40       & 3.80E15 & HC   & 2011/06/02 1100  & 830            & 5.78E-18                               & 2011/06/04 2045 & 480 & 510           & 34.3   & 14.9   & 7.2      & 5.1      \\
12(1) & 2011/06/14 0610     & $-$44                 & $-$1                  & 41                     & 1.00                & 0.28       & 1.03E16 & NS   & 2011/06/14 1300  & 770          & 2.14E-17                               & 2011/06/17 0241 & 530   & 490           & 5.6    & 7.0    & 57.3     & 51.5     \\
13(1) & 2011/08/04 0412     & 31                    & 20                    & 62                     & 0.87                & 0.49       & 6.84E15 & UG   & 2011/08/04 0550  & 1300          & 6.93E-18                               & 2011/08/05 1751 & 540 & 540          & 7.5    & 1.2    & 2.1      & 1.9      \\
14(1) & 2011/09/13 2210     & 19                    & 21                    & $-$6                   & 0.22                & 0.47       & 4.20E15 & HC   & 2011/09/14 0620  & 600           & 5.98E-18                               & 2011/09/17 0343 & 500   & 450         & 13.3   & 4.3    & 7.1      & 4.4      \\
15(1) & 2011/10/22 0005     & 87                    & 45                    & 16                     & 0.79                & 0.59       & 1.23E16 & NS   & 2011/10/22 1330  & 610          & 1.01E-17                               & 2011/10/24 1831 & 480  & 470         & 25.6   & 11.0   & 12.2     & 11.0     \\
16(2) & 2011/10/27 1200     & $-$37                 & 29                    & 17                     & 0.29                & 0.36       & 3.01E15 & NS   & 2011/10/27 1550  & 700            & 5.76E-18                               & 2011/11/01 0907 & 400 & 370             & 7.5    & 3.7    & 4.8      & 2.9      \\
17(1) & 2011/11/26 0700     & 56                    & 14                    & $-$41                  & 0.77                & 0.63       & 1.00E16 & HC   & 2011/11/26 1000  & 1350          & 7.67E-18                               & 2011/11/28 2150 & 500 & 450             & 9.2    & 13.4   & 22.3     & 20.1     \\
18(1) & 2012/01/19 1512     & $-$20                 & 44                    & 90                     & 1.00                & 0.47       & 9.17E15 & NS   & 2012/01/19 1730  & 1000       & 9.60E-18                               & 2012/01/22 0611 & 410 & 450              & 26.6   & 6.7    & 10.9     & 10.0     \\
19(2) & 2012/03/13 1736     & 62                    & 21                    & $-$40                  & 1.28                & 0.74       & 1.00E16 & NS   & 2012/03/13 1940  & 850   & 6.02E-18                               &   2012/03/15 1306 & 710 & 710            & 9.1    & 4.7    & 4.3      & 4.3      \\
20(2) & 2012/06/14 1412     & 1                     & $-$24                 & 67                     & 0.65                & 0.52       & 8.46E15 & UG   & 2012/06/14 1700  & 1000          & 8.39E-18                               & 2012/06/16 2019 & 490 & 450           & 37.7   & 22.6   & 15.7     & 13.2     \\
21(2) & 2012/07/12 1648     & 7                     & $-$18                 & 70                     & 0.39                & 0.59       & 1.84E16 & UG   & 2012/07/12 1900  & 1150           & 1.75E-17                               & 2012/07/14 1809 & 610 & 490            & 15.8   & 2.3    & 5.2      & 4.1      \\
22(2) & 2012/09/28 0000     & 11                    & 10                    & 75                     & 0.60                & 0.40       & 9.61E15 & UG   & 2012/09/28 0300  & 1000             & 1.34E-17                               & 2012/09/30 2305 & 370 & 370           & 20.3   & 10.8   & 19.8      & 15.5      \\
23(1) & 2012/10/05 0724    & 21                    & $-$18                 & 41                     & 0.57                & 0.40       & 6.40E15 & UG   & 2012/10/05 0900  & 600           & 9.05E-18                               & 2012/10/08 0516 & 370 & 400           & 15.7   & 5.2    & 8.3     & 6.4      \\
24(2) & 2012/11/09 1512     & $-$11                 & $-$18                 & 6                      & 0.60                & 0.48       & 5.19E15 & NS   & 2012/11/09 1900  & 600            & 5.77E-18                               & 2012/11/12 2204 & 410  & 380           & 22.3   & 6.5    & 6.5      & 5.2      \\
25(1) & 2012/11/23 1336     & $-$22                 & $-$21                 & $-$66                  & 0.18                & 0.56       & 3.45E15 & NS   & 2012/11/23 1810  & 680           & 4.06E-18                               & 2012/11/26 0512 & 520   & 450          & 7.3    & 2.6    & 2.1      & 1.4      \\
26(1) & 2013/04/11 0724     & $-$16                 & $-$3                  & 41                     & 0.52                & 0.40       & 1.51E16 & UG   & 2013/04/11 1130  & 700            & 2.18E-17                               & 2013/04/13 2254 & 490  & 410            & 13.4   & 4.1    & 11.2     & 8.3      \\
27(2) & 2013/07/09 1512     & 0                     & 2                     & 1                      & 0.50                & 0.38       & 2.93E15 & UG   & 2013/07/09 2150  & 550            & 4.57E-18                               & 2013/07/12 1714 & 480  & 410           & 5.7    & 3.7    & 2.1      & 1.5      \\
28(1) & 2013/09/29 2145     & 24                    & 23                    & 90                     & 0.82                & 0.43       & 1.37E16 & NS   & 2013/09/30 0100  & 1000          & 1.65E-17                               & 2013/10/02 0154 & 590  & 470         & 12.2   & 1.8    & 6.9      & 5.9      \\
29(2) & 2014/08/15 1812     & 15                    & 13                    & $-$65                  & 0.38                & 0.30       & 1.60E15 & UG   & 2014/08/16 0140  & 450           & 3.68E-18                               & 2014/08/19 0657 & 370  & 360         & 18.2    & 8.1    & 2.9      & 1.8     
\end{tabular}
\end{sidewaystable}

\acknowledgments
M.T., K.D. S.G.H., and A.M.V. acknowledge funding by the Austrian Space Applications Programme of the Austrian Research Promotion Agency FFG (ASAP-13 859729 SWAMI, ASAP-11 4900217 CORDIM and ASAP-14 865972 SSCME). M.D. acknowledges the support by H2020 MSCA project No 745782 (ForbMod). B.V. and M.D. acknowledge the support by the Croatian Science Foundation under the project No 7549 (MSOC). N.S. would like to acknowledge Prof. P Subramanian under whose supervision she finished her PhD thesis work at IISER Pune, India. C.S. is supported by the Research Foundation -- Flanders (FWO, grant no. 1S42817N). E.A. would like to acknowledge the financial support by the Finnish Academy of Science and Letters via the Postdoc Pool funding for the period October 2018--July 2019, and the Academy of Finland (Postdoctoral Grant No 322455) for the period September 2019 - September 2022. Data from the SOHO and STEREO mission are downloaded from the Virtual Solar Observatory Repository \cite{hill09}. In-situ plasma and interplanetary magnetic field measurements (5-minute averaged) are taken via OMNIweb from the Wind spacecraft and its Solar Wind Experiment \cite{ogilvie95} and Magnetic Field Experiment \cite{lepping95}.

 \bibliography{bib}

\begin{thebibliography}{}

\bibitem [\protect \citeauthoryear {%
{Bein}%
, {Temmer}%
, {Vourlidas}%
, {Veronig}%
\BCBL {}\ \BBA {} {Utz}%
}{%
{Bein}%
\ \protect \BOthers {.}}{%
{\protect \APACyear {2013}}%
}]{%
bein13}
\APACinsertmetastar {%
bein13}%
\begin{APACrefauthors}%
{Bein}, B\BPBI M.%
, {Temmer}, M.%
, {Vourlidas}, A.%
, {Veronig}, A\BPBI M.%
\BCBL {}\ \BBA {} {Utz}, D.%
\end{APACrefauthors}%
\unskip\
\newblock
\APACrefYearMonthDay{2013}{May}{}.
\newblock
{\BBOQ}\APACrefatitle {{The Height Evolution of the ``True'' Coronal Mass
  Ejection Mass derived from STEREO COR1 and COR2 Observations}} {{The Height
  Evolution of the ``True'' Coronal Mass Ejection Mass derived from STEREO COR1
  and COR2 Observations}}.{\BBCQ}
\newblock
\APACjournalVolNumPages{\apj}{768}{1}{31}.
\newblock
\begin{APACrefDOI} \doi{10.1088/0004-637X/768/1/31} \end{APACrefDOI}
\PrintBackRefs{\CurrentBib}

\bibitem [\protect \citeauthoryear {%
{Bemporad}%
\ \BBA {} {Mancuso}%
}{%
{Bemporad}%
\ \BBA {} {Mancuso}%
}{%
{\protect \APACyear {2010}}%
}]{%
bemporad10}
\APACinsertmetastar {%
bemporad10}%
\begin{APACrefauthors}%
{Bemporad}, A.%
\BCBT {}\ \BBA {} {Mancuso}, S.%
\end{APACrefauthors}%
\unskip\
\newblock
\APACrefYearMonthDay{2010}{Sep}{}.
\newblock
{\BBOQ}\APACrefatitle {{First Complete Determination of Plasma Physical
  Parameters Across a Coronal Mass Ejection-driven Shock}} {{First Complete
  Determination of Plasma Physical Parameters Across a Coronal Mass
  Ejection-driven Shock}}.{\BBCQ}
\newblock
\APACjournalVolNumPages{\apj}{720}{1}{130-143}.
\newblock
\begin{APACrefDOI} \doi{10.1088/0004-637X/720/1/130} \end{APACrefDOI}
\PrintBackRefs{\CurrentBib}

\bibitem [\protect \citeauthoryear {%
{Bothmer}%
\ \BBA {} {Schwenn}%
}{%
{Bothmer}%
\ \BBA {} {Schwenn}%
}{%
{\protect \APACyear {1998}}%
}]{%
1998bothmer+schwenn}
\APACinsertmetastar {%
1998bothmer+schwenn}%
\begin{APACrefauthors}%
{Bothmer}, V.%
\BCBT {}\ \BBA {} {Schwenn}, R.%
\end{APACrefauthors}%
\unskip\
\newblock
\APACrefYearMonthDay{1998}{{\APACmonth{01}}}{}.
\newblock
{\BBOQ}\APACrefatitle {{The structure and origin of magnetic clouds in the
  solar wind}} {{The structure and origin of magnetic clouds in the solar
  wind}}.{\BBCQ}
\newblock
\APACjournalVolNumPages{Annales Geophysicae}{16}{}{1-24}.
\newblock
\begin{APACrefDOI} \doi{10.1007/s00585-997-0001-x} \end{APACrefDOI}
\PrintBackRefs{\CurrentBib}

\bibitem [\protect \citeauthoryear {%
{Cargill}%
}{%
{Cargill}%
}{%
{\protect \APACyear {2004}}%
}]{%
cargill04}
\APACinsertmetastar {%
cargill04}%
\begin{APACrefauthors}%
{Cargill}, P\BPBI J.%
\end{APACrefauthors}%
\unskip\
\newblock
\APACrefYearMonthDay{2004}{{\APACmonth{05}}}{}.
\newblock
{\BBOQ}\APACrefatitle {{On the Aerodynamic Drag Force Acting on Interplanetary
  Coronal Mass Ejections}} {{On the Aerodynamic Drag Force Acting on
  Interplanetary Coronal Mass Ejections}}.{\BBCQ}
\newblock
\APACjournalVolNumPages{\solphys}{221}{}{135-149}.
\newblock
\begin{APACrefDOI} \doi{10.1023/B:SOLA.0000033366.10725.a2} \end{APACrefDOI}
\PrintBackRefs{\CurrentBib}

\bibitem [\protect \citeauthoryear {%
{Cargill}%
, {Chen}%
, {Spicer}%
\BCBL {}\ \BBA {} {Zalesak}%
}{%
{Cargill}%
\ \protect \BOthers {.}}{%
{\protect \APACyear {1996}}%
}]{%
cargill96}
\APACinsertmetastar {%
cargill96}%
\begin{APACrefauthors}%
{Cargill}, P\BPBI J.%
, {Chen}, J.%
, {Spicer}, D\BPBI S.%
\BCBL {}\ \BBA {} {Zalesak}, S\BPBI T.%
\end{APACrefauthors}%
\unskip\
\newblock
\APACrefYearMonthDay{1996}{{\APACmonth{03}}}{}.
\newblock
{\BBOQ}\APACrefatitle {{Magnetohydrodynamic simulations of the motion of
  magnetic flux tubes through a magnetized plasma}} {{Magnetohydrodynamic
  simulations of the motion of magnetic flux tubes through a magnetized
  plasma}}.{\BBCQ}
\newblock
\APACjournalVolNumPages{\jgr}{101}{}{4855-4870}.
\newblock
\begin{APACrefDOI} \doi{10.1029/95JA03769} \end{APACrefDOI}
\PrintBackRefs{\CurrentBib}

\bibitem [\protect \citeauthoryear {%
{Chen}%
}{%
{Chen}%
}{%
{\protect \APACyear {1989}}%
}]{%
chen89}
\APACinsertmetastar {%
chen89}%
\begin{APACrefauthors}%
{Chen}, J.%
\end{APACrefauthors}%
\unskip\
\newblock
\APACrefYearMonthDay{1989}{{\APACmonth{03}}}{}.
\newblock
{\BBOQ}\APACrefatitle {{Effects of toroidal forces in current loops embedded in
  a background plasma}} {{Effects of toroidal forces in current loops embedded
  in a background plasma}}.{\BBCQ}
\newblock
\APACjournalVolNumPages{\apj}{338}{}{453-470}.
\newblock
\begin{APACrefDOI} \doi{10.1086/167211} \end{APACrefDOI}
\PrintBackRefs{\CurrentBib}

\bibitem [\protect \citeauthoryear {%
{Chen}%
}{%
{Chen}%
}{%
{\protect \APACyear {1996}}%
}]{%
chen96}
\APACinsertmetastar {%
chen96}%
\begin{APACrefauthors}%
{Chen}, J.%
\end{APACrefauthors}%
\unskip\
\newblock
\APACrefYearMonthDay{1996}{{\APACmonth{12}}}{}.
\newblock
{\BBOQ}\APACrefatitle {{Theory of prominence eruption and propagation:
  Interplanetary consequences}} {{Theory of prominence eruption and
  propagation: Interplanetary consequences}}.{\BBCQ}
\newblock
\APACjournalVolNumPages{\jgr}{101}{}{27499-27520}.
\newblock
\begin{APACrefDOI} \doi{10.1029/96JA02644} \end{APACrefDOI}
\PrintBackRefs{\CurrentBib}

\bibitem [\protect \citeauthoryear {%
{Colaninno}%
\ \BBA {} {Vourlidas}%
}{%
{Colaninno}%
\ \BBA {} {Vourlidas}%
}{%
{\protect \APACyear {2009}}%
}]{%
colaninno09}
\APACinsertmetastar {%
colaninno09}%
\begin{APACrefauthors}%
{Colaninno}, R\BPBI C.%
\BCBT {}\ \BBA {} {Vourlidas}, A.%
\end{APACrefauthors}%
\unskip\
\newblock
\APACrefYearMonthDay{2009}{Jun}{}.
\newblock
{\BBOQ}\APACrefatitle {{First Determination of the True Mass of Coronal Mass
  Ejections: A Novel Approach to Using the Two STEREO Viewpoints}} {{First
  Determination of the True Mass of Coronal Mass Ejections: A Novel Approach to
  Using the Two STEREO Viewpoints}}.{\BBCQ}
\newblock
\APACjournalVolNumPages{\apj}{698}{1}{852-858}.
\newblock
\begin{APACrefDOI} \doi{10.1088/0004-637X/698/1/852} \end{APACrefDOI}
\PrintBackRefs{\CurrentBib}

\bibitem [\protect \citeauthoryear {%
{Dasso}%
, {Nakwacki}%
, {D{\'e}moulin}%
\BCBL {}\ \BBA {} {Mand rini}%
}{%
{Dasso}%
\ \protect \BOthers {.}}{%
{\protect \APACyear {2007}}%
}]{%
dasso07}
\APACinsertmetastar {%
dasso07}%
\begin{APACrefauthors}%
{Dasso}, S.%
, {Nakwacki}, M\BPBI S.%
, {D{\'e}moulin}, P.%
\BCBL {}\ \BBA {} {Mand rini}, C\BPBI H.%
\end{APACrefauthors}%
\unskip\
\newblock
\APACrefYearMonthDay{2007}{Aug}{}.
\newblock
{\BBOQ}\APACrefatitle {{Progressive Transformation of a Flux Rope to an ICME.
  Comparative Analysis Using the Direct and Fitted Expansion Methods}}
  {{Progressive Transformation of a Flux Rope to an ICME. Comparative Analysis
  Using the Direct and Fitted Expansion Methods}}.{\BBCQ}
\newblock
\APACjournalVolNumPages{\solphys}{244}{1-2}{115-137}.
\newblock
\begin{APACrefDOI} \doi{10.1007/s11207-007-9034-2} \end{APACrefDOI}
\PrintBackRefs{\CurrentBib}

\bibitem [\protect \citeauthoryear {%
{DeForest}%
, {Howard}%
\BCBL {}\ \BBA {} {McComas}%
}{%
{DeForest}%
\ \protect \BOthers {.}}{%
{\protect \APACyear {2013}}%
}]{%
deforest13}
\APACinsertmetastar {%
deforest13}%
\begin{APACrefauthors}%
{DeForest}, C\BPBI E.%
, {Howard}, T\BPBI A.%
\BCBL {}\ \BBA {} {McComas}, D\BPBI J.%
\end{APACrefauthors}%
\unskip\
\newblock
\APACrefYearMonthDay{2013}{May}{}.
\newblock
{\BBOQ}\APACrefatitle {{Tracking Coronal Features from the Low Corona to Earth:
  A Quantitative Analysis of the 2008 December 12 Coronal Mass Ejection}}
  {{Tracking Coronal Features from the Low Corona to Earth: A Quantitative
  Analysis of the 2008 December 12 Coronal Mass Ejection}}.{\BBCQ}
\newblock
\APACjournalVolNumPages{\apj}{769}{1}{43}.
\newblock
\begin{APACrefDOI} \doi{10.1088/0004-637X/769/1/43} \end{APACrefDOI}
\PrintBackRefs{\CurrentBib}

\bibitem [\protect \citeauthoryear {%
{D{\'e}moulin}%
, {Nakwacki}%
, {Dasso}%
\BCBL {}\ \BBA {} {Mandrini}%
}{%
{D{\'e}moulin}%
\ \protect \BOthers {.}}{%
{\protect \APACyear {2008}}%
}]{%
2008demoulin_solphys}
\APACinsertmetastar {%
2008demoulin_solphys}%
\begin{APACrefauthors}%
{D{\'e}moulin}, P.%
, {Nakwacki}, M\BPBI S.%
, {Dasso}, S.%
\BCBL {}\ \BBA {} {Mandrini}, C\BPBI H.%
\end{APACrefauthors}%
\unskip\
\newblock
\APACrefYearMonthDay{2008}{{\APACmonth{08}}}{}.
\newblock
{\BBOQ}\APACrefatitle {{Expected in Situ Velocities from a Hierarchical Model
  for Expanding Interplanetary Coronal Mass Ejections}} {{Expected in Situ
  Velocities from a Hierarchical Model for Expanding Interplanetary Coronal
  Mass Ejections}}.{\BBCQ}
\newblock
\APACjournalVolNumPages{\solphys}{250}{}{347-374}.
\newblock
\begin{APACrefDOI} \doi{10.1007/s11207-008-9221-9} \end{APACrefDOI}
\PrintBackRefs{\CurrentBib}

\bibitem [\protect \citeauthoryear {%
{Domingo}%
, {Fleck}%
\BCBL {}\ \BBA {} {Poland}%
}{%
{Domingo}%
\ \protect \BOthers {.}}{%
{\protect \APACyear {1995}}%
}]{%
1995soho}
\APACinsertmetastar {%
1995soho}%
\begin{APACrefauthors}%
{Domingo}, V.%
, {Fleck}, B.%
\BCBL {}\ \BBA {} {Poland}, A\BPBI I.%
\end{APACrefauthors}%
\unskip\
\newblock
\APACrefYearMonthDay{1995}{{\APACmonth{12}}}{}.
\newblock
{\BBOQ}\APACrefatitle {{The SOHO Mission: an Overview}} {{The SOHO Mission: an
  Overview}}.{\BBCQ}
\newblock
\APACjournalVolNumPages{\solphys}{162}{}{1-37}.
\newblock
\begin{APACrefDOI} \doi{10.1007/BF00733425} \end{APACrefDOI}
\PrintBackRefs{\CurrentBib}

\bibitem [\protect \citeauthoryear {%
{Dumbovi{\'c}}%
\ \protect \BOthers {.}}{%
{Dumbovi{\'c}}%
\ \protect \BOthers {.}}{%
{\protect \APACyear {2019}}%
}]{%
dumbovic19}
\APACinsertmetastar {%
dumbovic19}%
\begin{APACrefauthors}%
{Dumbovi{\'c}}, M.%
, {Guo}, J.%
, {Temmer}, M.%
, {Mays}, M\BPBI L.%
, {Veronig}, A.%
, {Heinemann}, S\BPBI G.%
\BDBL {}{Leitner}, M.%
\end{APACrefauthors}%
\unskip\
\newblock
\APACrefYearMonthDay{2019}{{\APACmonth{07}}}{}.
\newblock
{\BBOQ}\APACrefatitle {{Unusual Plasma and Particle Signatures at Mars and
  STEREO-A Related to CME-CME Interaction}} {{Unusual Plasma and Particle
  Signatures at Mars and STEREO-A Related to CME-CME Interaction}}.{\BBCQ}
\newblock
\APACjournalVolNumPages{\apj}{880}{1}{18}.
\newblock
\begin{APACrefDOI} \doi{10.3847/1538-4357/ab27ca} \end{APACrefDOI}
\PrintBackRefs{\CurrentBib}

\bibitem [\protect \citeauthoryear {%
{Dumbovi{\'c}}%
\ \protect \BOthers {.}}{%
{Dumbovi{\'c}}%
\ \protect \BOthers {.}}{%
{\protect \APACyear {2018}}%
}]{%
dumbovic18}
\APACinsertmetastar {%
dumbovic18}%
\begin{APACrefauthors}%
{Dumbovi{\'c}}, M.%
, {{\v{C}}alogovi{\'c}}, J.%
, {Vr{\v{s}}nak}, B.%
, {Temmer}, M.%
, {Mays}, M\BPBI L.%
, {Veronig}, A.%
\BCBL {}\ \BBA {} {Piantschitsch}, I.%
\end{APACrefauthors}%
\unskip\
\newblock
\APACrefYearMonthDay{2018}{Feb}{}.
\newblock
{\BBOQ}\APACrefatitle {{The Drag-based Ensemble Model (DBEM) for Coronal Mass
  Ejection Propagation}} {{The Drag-based Ensemble Model (DBEM) for Coronal
  Mass Ejection Propagation}}.{\BBCQ}
\newblock
\APACjournalVolNumPages{\apj}{854}{2}{180}.
\newblock
\begin{APACrefDOI} \doi{10.3847/1538-4357/aaaa66} \end{APACrefDOI}
\PrintBackRefs{\CurrentBib}

\bibitem [\protect \citeauthoryear {%
Efron%
}{%
Efron%
}{%
{\protect \APACyear {1979}}%
}]{%
efron1979_bootstrap}
\APACinsertmetastar {%
efron1979_bootstrap}%
\begin{APACrefauthors}%
Efron, B.%
\end{APACrefauthors}%
\unskip\
\newblock
\APACrefYearMonthDay{1979}{01}{}.
\newblock
{\BBOQ}\APACrefatitle {Bootstrap Methods: Another Look at the Jackknife}
  {Bootstrap methods: Another look at the jackknife}.{\BBCQ}
\newblock
\APACjournalVolNumPages{Ann. Statist.}{7}{1}{1--26}.
\newblock
\begin{APACrefURL} \url{https://doi.org/10.1214/aos/1176344552}
  \end{APACrefURL}
\newblock
\begin{APACrefDOI} \doi{10.1214/aos/1176344552} \end{APACrefDOI}
\PrintBackRefs{\CurrentBib}

\bibitem [\protect \citeauthoryear {%
Efron%
\ \BBA {} Tibshirani%
}{%
Efron%
\ \BBA {} Tibshirani%
}{%
{\protect \APACyear {1993}}%
}]{%
efron93_bootstrap}
\APACinsertmetastar {%
efron93_bootstrap}%
\begin{APACrefauthors}%
Efron, B.%
\BCBT {}\ \BBA {} Tibshirani, R\BPBI J.%
\end{APACrefauthors}%
\unskip\
\newblock
\APACrefYear{1993}.
\newblock
\APACrefbtitle {{An Introduction to the Bootstrap}} {{An Introduction to the
  Bootstrap}}.
\newblock
\APACaddressPublisher{New York}{Chapman \& Hall}.
\PrintBackRefs{\CurrentBib}

\bibitem [\protect \citeauthoryear {%
{Fox}%
\ \protect \BOthers {.}}{%
{Fox}%
\ \protect \BOthers {.}}{%
{\protect \APACyear {2016}}%
}]{%
fox16}
\APACinsertmetastar {%
fox16}%
\begin{APACrefauthors}%
{Fox}, N\BPBI J.%
, {Velli}, M\BPBI C.%
, {Bale}, S\BPBI D.%
, {Decker}, R.%
, {Driesman}, A.%
, {Howard}, R\BPBI A.%
\BDBL {}{Szabo}, A.%
\end{APACrefauthors}%
\unskip\
\newblock
\APACrefYearMonthDay{2016}{{\APACmonth{12}}}{}.
\newblock
{\BBOQ}\APACrefatitle {{The Solar Probe Plus Mission: Humanity's First Visit to
  Our Star}} {{The Solar Probe Plus Mission: Humanity's First Visit to Our
  Star}}.{\BBCQ}
\newblock
\APACjournalVolNumPages{\ssr}{204}{1-4}{7-48}.
\newblock
\begin{APACrefDOI} \doi{10.1007/s11214-015-0211-6} \end{APACrefDOI}
\PrintBackRefs{\CurrentBib}

\bibitem [\protect \citeauthoryear {%
{Gonzalez}%
\ \protect \BOthers {.}}{%
{Gonzalez}%
\ \protect \BOthers {.}}{%
{\protect \APACyear {1989}}%
}]{%
gonzalez89}
\APACinsertmetastar {%
gonzalez89}%
\begin{APACrefauthors}%
{Gonzalez}, W\BPBI D.%
, {Tsurutani}, B\BPBI T.%
, {Gonzalez}, A\BPBI L\BPBI C.%
, {Smith}, E\BPBI J.%
, {Tang}, F.%
\BCBL {}\ \BBA {} {Akasofu}, S\BHBI I.%
\end{APACrefauthors}%
\unskip\
\newblock
\APACrefYearMonthDay{1989}{Jul}{}.
\newblock
{\BBOQ}\APACrefatitle {{Solar wind-magnetosphere coupling during intense
  magnetic storms (1978-1979)}} {{Solar wind-magnetosphere coupling during
  intense magnetic storms (1978-1979)}}.{\BBCQ}
\newblock
\APACjournalVolNumPages{\jgr}{94}{A7}{8835-8851}.
\newblock
\begin{APACrefDOI} \doi{10.1029/JA094iA07p08835} \end{APACrefDOI}
\PrintBackRefs{\CurrentBib}

\bibitem [\protect \citeauthoryear {%
{Gulisano}%
, {D{\'e}moulin}%
, {Dasso}%
\BCBL {}\ \BBA {} {Rodriguez}%
}{%
{Gulisano}%
\ \protect \BOthers {.}}{%
{\protect \APACyear {2012}}%
}]{%
2012gulisano}
\APACinsertmetastar {%
2012gulisano}%
\begin{APACrefauthors}%
{Gulisano}, A\BPBI M.%
, {D{\'e}moulin}, P.%
, {Dasso}, S.%
\BCBL {}\ \BBA {} {Rodriguez}, L.%
\end{APACrefauthors}%
\unskip\
\newblock
\APACrefYearMonthDay{2012}{{\APACmonth{07}}}{}.
\newblock
{\BBOQ}\APACrefatitle {{Expansion of magnetic clouds in the outer heliosphere}}
  {{Expansion of magnetic clouds in the outer heliosphere}}.{\BBCQ}
\newblock
\APACjournalVolNumPages{\aap}{543}{}{A107}.
\newblock
\begin{APACrefDOI} \doi{10.1051/0004-6361/201118748} \end{APACrefDOI}
\PrintBackRefs{\CurrentBib}

\bibitem [\protect \citeauthoryear {%
{Hill}%
\ \protect \BOthers {.}}{%
{Hill}%
\ \protect \BOthers {.}}{%
{\protect \APACyear {2009}}%
}]{%
hill09}
\APACinsertmetastar {%
hill09}%
\begin{APACrefauthors}%
{Hill}, F.%
, {Martens}, P.%
, {Yoshimura}, K.%
, {Gurman}, J.%
, {Hourcl{\'e}}, J.%
, {Dimitoglou}, G.%
\BDBL {}{Tian}, K\BPBI Q.%
\end{APACrefauthors}%
\unskip\
\newblock
\APACrefYearMonthDay{2009}{{\APACmonth{04}}}{}.
\newblock
{\BBOQ}\APACrefatitle {{The Virtual Solar Observatory{\textemdash}A Resource
  for International Heliophysics Research}} {{The Virtual Solar
  Observatory{\textemdash}A Resource for International Heliophysics
  Research}}.{\BBCQ}
\newblock
\APACjournalVolNumPages{Earth Moon and Planets}{104}{1-4}{315-330}.
\newblock
\begin{APACrefDOI} \doi{10.1007/s11038-008-9274-7} \end{APACrefDOI}
\PrintBackRefs{\CurrentBib}

\bibitem [\protect \citeauthoryear {%
{Holzknecht}%
\ \protect \BOthers {.}}{%
{Holzknecht}%
\ \protect \BOthers {.}}{%
{\protect \APACyear {2018}}%
}]{%
holzknecht18}
\APACinsertmetastar {%
holzknecht18}%
\begin{APACrefauthors}%
{Holzknecht}, L.%
, {Temmer}, M.%
, {Dumbovi{\'c}}, M.%
, {Wellenzohn}, S.%
, {Krikova}, K.%
, {Heinemann}, S\BPBI G.%
\BDBL {}{Veronig}, A\BPBI M.%
\end{APACrefauthors}%
\unskip\
\newblock
\APACrefYearMonthDay{2018}{{\APACmonth{01}}}{}.
\newblock
{\BBOQ}\APACrefatitle {{CME volume calculation from 3D GCS reconstruction}}
  {{CME volume calculation from 3D GCS reconstruction}}.{\BBCQ}
\newblock
\APACjournalVolNumPages{Central European Astrophysical Bulletin}{42}{}{3}.
\PrintBackRefs{\CurrentBib}

\bibitem [\protect \citeauthoryear {%
{Howard}%
\ \protect \BOthers {.}}{%
{Howard}%
\ \protect \BOthers {.}}{%
{\protect \APACyear {2008}}%
}]{%
2008howard_SECCHI}
\APACinsertmetastar {%
2008howard_SECCHI}%
\begin{APACrefauthors}%
{Howard}, R\BPBI A.%
, {Moses}, J\BPBI D.%
, {Vourlidas}, A.%
, {Newmark}, J\BPBI S.%
, {Socker}, D\BPBI G.%
, {Plunkett}, S\BPBI P.%
\BDBL {}{Carter}, T.%
\end{APACrefauthors}%
\unskip\
\newblock
\APACrefYearMonthDay{2008}{{\APACmonth{04}}}{}.
\newblock
{\BBOQ}\APACrefatitle {{Sun Earth Connection Coronal and Heliospheric
  Investigation (SECCHI)}} {{Sun Earth Connection Coronal and Heliospheric
  Investigation (SECCHI)}}.{\BBCQ}
\newblock
\APACjournalVolNumPages{\ssr}{136}{}{67-115}.
\newblock
\begin{APACrefDOI} \doi{10.1007/s11214-008-9341-4} \end{APACrefDOI}
\PrintBackRefs{\CurrentBib}

\bibitem [\protect \citeauthoryear {%
{Howard}%
\ \BBA {} {Vourlidas}%
}{%
{Howard}%
\ \BBA {} {Vourlidas}%
}{%
{\protect \APACyear {2018}}%
}]{%
howard18}
\APACinsertmetastar {%
howard18}%
\begin{APACrefauthors}%
{Howard}, R\BPBI A.%
\BCBT {}\ \BBA {} {Vourlidas}, A.%
\end{APACrefauthors}%
\unskip\
\newblock
\APACrefYearMonthDay{2018}{Apr}{}.
\newblock
{\BBOQ}\APACrefatitle {{Evolution of CME Mass in the Corona}} {{Evolution of
  CME Mass in the Corona}}.{\BBCQ}
\newblock
\APACjournalVolNumPages{\solphys}{293}{4}{55}.
\newblock
\begin{APACrefDOI} \doi{10.1007/s11207-018-1274-9} \end{APACrefDOI}
\PrintBackRefs{\CurrentBib}

\bibitem [\protect \citeauthoryear {%
{Janvier}%
\ \protect \BOthers {.}}{%
{Janvier}%
\ \protect \BOthers {.}}{%
{\protect \APACyear {2019}}%
}]{%
janvier19}
\APACinsertmetastar {%
janvier19}%
\begin{APACrefauthors}%
{Janvier}, M.%
, {Winslow}, R\BPBI M.%
, {Good}, S.%
, {Bonhomme}, E.%
, {D{\'e}moulin}, P.%
, {Dasso}, S.%
\BDBL {}{Boakes}, P\BPBI D.%
\end{APACrefauthors}%
\unskip\
\newblock
\APACrefYearMonthDay{2019}{Feb}{}.
\newblock
{\BBOQ}\APACrefatitle {{Generic Magnetic Field Intensity Profiles of
  Interplanetary Coronal Mass Ejections at Mercury, Venus, and Earth From
  Superposed Epoch Analyses}} {{Generic Magnetic Field Intensity Profiles of
  Interplanetary Coronal Mass Ejections at Mercury, Venus, and Earth From
  Superposed Epoch Analyses}}.{\BBCQ}
\newblock
\APACjournalVolNumPages{Journal of Geophysical Research (Space
  Physics)}{124}{2}{812-836}.
\newblock
\begin{APACrefDOI} \doi{10.1029/2018JA025949} \end{APACrefDOI}
\PrintBackRefs{\CurrentBib}

\bibitem [\protect \citeauthoryear {%
{Kaiser}%
\ \protect \BOthers {.}}{%
{Kaiser}%
\ \protect \BOthers {.}}{%
{\protect \APACyear {2008}}%
}]{%
kaiser08}
\APACinsertmetastar {%
kaiser08}%
\begin{APACrefauthors}%
{Kaiser}, M\BPBI L.%
, {Kucera}, T\BPBI A.%
, {Davila}, J\BPBI M.%
, {St. Cyr}, O\BPBI C.%
, {Guhathakurta}, M.%
\BCBL {}\ \BBA {} {Christian}, E.%
\end{APACrefauthors}%
\unskip\
\newblock
\APACrefYearMonthDay{2008}{Apr}{}.
\newblock
{\BBOQ}\APACrefatitle {{The STEREO Mission: An Introduction}} {{The STEREO
  Mission: An Introduction}}.{\BBCQ}
\newblock
\APACjournalVolNumPages{\ssr}{136}{1-4}{5-16}.
\newblock
\begin{APACrefDOI} \doi{10.1007/s11214-007-9277-0} \end{APACrefDOI}
\PrintBackRefs{\CurrentBib}

\bibitem [\protect \citeauthoryear {%
{Kilpua}%
, {Koskinen}%
\BCBL {}\ \BBA {} {Pulkkinen}%
}{%
{Kilpua}%
\ \protect \BOthers {.}}{%
{\protect \APACyear {2017}}%
}]{%
kilpua17b}
\APACinsertmetastar {%
kilpua17b}%
\begin{APACrefauthors}%
{Kilpua}, E.%
, {Koskinen}, H\BPBI E\BPBI J.%
\BCBL {}\ \BBA {} {Pulkkinen}, T\BPBI I.%
\end{APACrefauthors}%
\unskip\
\newblock
\APACrefYearMonthDay{2017}{Nov}{}.
\newblock
{\BBOQ}\APACrefatitle {{Coronal mass ejections and their sheath regions in
  interplanetary space}} {{Coronal mass ejections and their sheath regions in
  interplanetary space}}.{\BBCQ}
\newblock
\APACjournalVolNumPages{Living Reviews in Solar Physics}{14}{1}{5}.
\newblock
\begin{APACrefDOI} \doi{10.1007/s41116-017-0009-6} \end{APACrefDOI}
\PrintBackRefs{\CurrentBib}

\bibitem [\protect \citeauthoryear {%
{Kliem}%
\ \BBA {} {T{\"o}r{\"o}k}%
}{%
{Kliem}%
\ \BBA {} {T{\"o}r{\"o}k}%
}{%
{\protect \APACyear {2006}}%
}]{%
kliem06}
\APACinsertmetastar {%
kliem06}%
\begin{APACrefauthors}%
{Kliem}, B.%
\BCBT {}\ \BBA {} {T{\"o}r{\"o}k}, T.%
\end{APACrefauthors}%
\unskip\
\newblock
\APACrefYearMonthDay{2006}{{\APACmonth{06}}}{}.
\newblock
{\BBOQ}\APACrefatitle {{Torus Instability}} {{Torus Instability}}.{\BBCQ}
\newblock
\APACjournalVolNumPages{Physical Review Letters}{96}{25}{255002-+}.
\newblock
\begin{APACrefDOI} \doi{10.1103/PhysRevLett.96.255002} \end{APACrefDOI}
\PrintBackRefs{\CurrentBib}

\bibitem [\protect \citeauthoryear {%
{Leblanc}%
, {Dulk}%
\BCBL {}\ \BBA {} {Bougeret}%
}{%
{Leblanc}%
\ \protect \BOthers {.}}{%
{\protect \APACyear {1998}}%
}]{%
leblanc98}
\APACinsertmetastar {%
leblanc98}%
\begin{APACrefauthors}%
{Leblanc}, Y.%
, {Dulk}, G\BPBI A.%
\BCBL {}\ \BBA {} {Bougeret}, J\BHBI L.%
\end{APACrefauthors}%
\unskip\
\newblock
\APACrefYearMonthDay{1998}{Nov}{}.
\newblock
{\BBOQ}\APACrefatitle {{Tracing the Electron Density from the Corona to 1au}}
  {{Tracing the Electron Density from the Corona to 1au}}.{\BBCQ}
\newblock
\APACjournalVolNumPages{\solphys}{183}{1}{165-180}.
\newblock
\begin{APACrefDOI} \doi{10.1023/A:1005049730506} \end{APACrefDOI}
\PrintBackRefs{\CurrentBib}

\bibitem [\protect \citeauthoryear {%
{Leitner}%
\ \protect \BOthers {.}}{%
{Leitner}%
\ \protect \BOthers {.}}{%
{\protect \APACyear {2007}}%
}]{%
2007leitner}
\APACinsertmetastar {%
2007leitner}%
\begin{APACrefauthors}%
{Leitner}, M.%
, {Farrugia}, C\BPBI J.%
, {M{\"o}Stl}, C.%
, {Ogilvie}, K\BPBI W.%
, {Galvin}, A\BPBI B.%
, {Schwenn}, R.%
\BCBL {}\ \BBA {} {Biernat}, H\BPBI K.%
\end{APACrefauthors}%
\unskip\
\newblock
\APACrefYearMonthDay{2007}{{\APACmonth{06}}}{}.
\newblock
{\BBOQ}\APACrefatitle {{Consequences of the force-free model of magnetic clouds
  for their heliospheric evolution}} {{Consequences of the force-free model of
  magnetic clouds for their heliospheric evolution}}.{\BBCQ}
\newblock
\APACjournalVolNumPages{Journal of Geophysical Research (Space
  Physics)}{112}{}{A06113}.
\newblock
\begin{APACrefDOI} \doi{10.1029/2006JA011940} \end{APACrefDOI}
\PrintBackRefs{\CurrentBib}

\bibitem [\protect \citeauthoryear {%
{Lepping}%
\ \protect \BOthers {.}}{%
{Lepping}%
\ \protect \BOthers {.}}{%
{\protect \APACyear {1995}}%
}]{%
lepping95}
\APACinsertmetastar {%
lepping95}%
\begin{APACrefauthors}%
{Lepping}, R\BPBI P.%
, {Ac{\~{u}}na}, M\BPBI H.%
, {Burlaga}, L\BPBI F.%
, {Farrell}, W\BPBI M.%
, {Slavin}, J\BPBI A.%
, {Schatten}, K\BPBI H.%
\BDBL {}{Worley}, E\BPBI M.%
\end{APACrefauthors}%
\unskip\
\newblock
\APACrefYearMonthDay{1995}{Feb}{}.
\newblock
{\BBOQ}\APACrefatitle {{The Wind Magnetic Field Investigation}} {{The Wind
  Magnetic Field Investigation}}.{\BBCQ}
\newblock
\APACjournalVolNumPages{\ssr}{71}{1-4}{207-229}.
\newblock
\begin{APACrefDOI} \doi{10.1007/BF00751330} \end{APACrefDOI}
\PrintBackRefs{\CurrentBib}

\bibitem [\protect \citeauthoryear {%
{Manchester}%
, {Kozyra}%
, {Lepri}%
\BCBL {}\ \BBA {} {Lavraud}%
}{%
{Manchester}%
\ \protect \BOthers {.}}{%
{\protect \APACyear {2014}}%
}]{%
manchester14}
\APACinsertmetastar {%
manchester14}%
\begin{APACrefauthors}%
{Manchester}, W\BPBI B.%
, {Kozyra}, J\BPBI U.%
, {Lepri}, S\BPBI T.%
\BCBL {}\ \BBA {} {Lavraud}, B.%
\end{APACrefauthors}%
\unskip\
\newblock
\APACrefYearMonthDay{2014}{Jul}{}.
\newblock
{\BBOQ}\APACrefatitle {{Simulation of magnetic cloud erosion during
  propagation}} {{Simulation of magnetic cloud erosion during
  propagation}}.{\BBCQ}
\newblock
\APACjournalVolNumPages{Journal of Geophysical Research (Space
  Physics)}{119}{7}{5449-5464}.
\newblock
\begin{APACrefDOI} \doi{10.1002/2014JA019882} \end{APACrefDOI}
\PrintBackRefs{\CurrentBib}

\bibitem [\protect \citeauthoryear {%
{Mas{\'{\i}}as-Meza}%
, {Dasso}%
, {D{\'e}moulin}%
, {Rodriguez}%
\BCBL {}\ \BBA {} {Janvier}%
}{%
{Mas{\'{\i}}as-Meza}%
\ \protect \BOthers {.}}{%
{\protect \APACyear {2016}}%
}]{%
masias16}
\APACinsertmetastar {%
masias16}%
\begin{APACrefauthors}%
{Mas{\'{\i}}as-Meza}, J\BPBI J.%
, {Dasso}, S.%
, {D{\'e}moulin}, P.%
, {Rodriguez}, L.%
\BCBL {}\ \BBA {} {Janvier}, M.%
\end{APACrefauthors}%
\unskip\
\newblock
\APACrefYearMonthDay{2016}{{\APACmonth{08}}}{}.
\newblock
{\BBOQ}\APACrefatitle {{Superposed epoch study of ICME sub-structures near
  Earth and their effects on Galactic cosmic rays}} {{Superposed epoch study of
  ICME sub-structures near Earth and their effects on Galactic cosmic
  rays}}.{\BBCQ}
\newblock
\APACjournalVolNumPages{\aap}{592}{}{A118}.
\newblock
\begin{APACrefDOI} \doi{10.1051/0004-6361/201628571} \end{APACrefDOI}
\PrintBackRefs{\CurrentBib}

\bibitem [\protect \citeauthoryear {%
{Mays}%
\ \protect \BOthers {.}}{%
{Mays}%
\ \protect \BOthers {.}}{%
{\protect \APACyear {2015}}%
}]{%
mays15}
\APACinsertmetastar {%
mays15}%
\begin{APACrefauthors}%
{Mays}, M\BPBI L.%
, {Taktakishvili}, A.%
, {Pulkkinen}, A.%
, {MacNeice}, P\BPBI J.%
, {Rast{\"a}tter}, L.%
, {Odstrcil}, D.%
\BDBL {}{Kuznetsova}, M\BPBI M.%
\end{APACrefauthors}%
\unskip\
\newblock
\APACrefYearMonthDay{2015}{Jun}{}.
\newblock
{\BBOQ}\APACrefatitle {{Ensemble Modeling of CMEs Using the WSA-ENLIL+Cone
  Model}} {{Ensemble Modeling of CMEs Using the WSA-ENLIL+Cone Model}}.{\BBCQ}
\newblock
\APACjournalVolNumPages{\solphys}{290}{6}{1775-1814}.
\newblock
\begin{APACrefDOI} \doi{10.1007/s11207-015-0692-1} \end{APACrefDOI}
\PrintBackRefs{\CurrentBib}

\bibitem [\protect \citeauthoryear {%
{M{\"u}ller}%
, {Zouganelis}%
, {St. Cyr}%
, {Gilbert}%
\BCBL {}\ \BBA {} {Nieves-Chinchilla}%
}{%
{M{\"u}ller}%
\ \protect \BOthers {.}}{%
{\protect \APACyear {2020}}%
}]{%
mueller20}
\APACinsertmetastar {%
mueller20}%
\begin{APACrefauthors}%
{M{\"u}ller}, D.%
, {Zouganelis}, I.%
, {St. Cyr}, O\BPBI C.%
, {Gilbert}, H\BPBI R.%
\BCBL {}\ \BBA {} {Nieves-Chinchilla}, T.%
\end{APACrefauthors}%
\unskip\
\newblock
\APACrefYearMonthDay{2020}{{\APACmonth{02}}}{}.
\newblock
{\BBOQ}\APACrefatitle {{Europe's next mission to the Sun}} {{Europe's next
  mission to the Sun}}.{\BBCQ}
\newblock
\APACjournalVolNumPages{Nature Astronomy}{4}{}{205-205}.
\newblock
\begin{APACrefDOI} \doi{10.1038/s41550-020-1015-5} \end{APACrefDOI}
\PrintBackRefs{\CurrentBib}

\bibitem [\protect \citeauthoryear {%
{Odstr{\v c}il}%
\ \BBA {} {Pizzo}%
}{%
{Odstr{\v c}il}%
\ \BBA {} {Pizzo}%
}{%
{\protect \APACyear {1999}}%
}]{%
odstrcil99}
\APACinsertmetastar {%
odstrcil99}%
\begin{APACrefauthors}%
{Odstr{\v c}il}, D.%
\BCBT {}\ \BBA {} {Pizzo}, V\BPBI J.%
\end{APACrefauthors}%
\unskip\
\newblock
\APACrefYearMonthDay{1999}{{\APACmonth{01}}}{}.
\newblock
{\BBOQ}\APACrefatitle {{Three-dimensional propagation of CMEs in a structured
  solar wind flow: 1. CME launched within the streamer belt}}
  {{Three-dimensional propagation of CMEs in a structured solar wind flow: 1.
  CME launched within the streamer belt}}.{\BBCQ}
\newblock
\APACjournalVolNumPages{J. Geophys. Res.}{104}{}{483-492}.
\newblock
\begin{APACrefDOI} \doi{10.1029/1998JA900019} \end{APACrefDOI}
\PrintBackRefs{\CurrentBib}

\bibitem [\protect \citeauthoryear {%
{Ogilvie}%
\ \protect \BOthers {.}}{%
{Ogilvie}%
\ \protect \BOthers {.}}{%
{\protect \APACyear {1995}}%
}]{%
ogilvie95}
\APACinsertmetastar {%
ogilvie95}%
\begin{APACrefauthors}%
{Ogilvie}, K\BPBI W.%
, {Chornay}, D\BPBI J.%
, {Fritzenreiter}, R\BPBI J.%
, {Hunsaker}, F.%
, {Keller}, J.%
, {Lobell}, J.%
\BDBL {}{Gergin}, E.%
\end{APACrefauthors}%
\unskip\
\newblock
\APACrefYearMonthDay{1995}{Feb}{}.
\newblock
{\BBOQ}\APACrefatitle {{SWE, A Comprehensive Plasma Instrument for the Wind
  Spacecraft}} {{SWE, A Comprehensive Plasma Instrument for the Wind
  Spacecraft}}.{\BBCQ}
\newblock
\APACjournalVolNumPages{\ssr}{71}{1-4}{55-77}.
\newblock
\begin{APACrefDOI} \doi{10.1007/BF00751326} \end{APACrefDOI}
\PrintBackRefs{\CurrentBib}

\bibitem [\protect \citeauthoryear {%
{Ontiveros}%
\ \BBA {} {Vourlidas}%
}{%
{Ontiveros}%
\ \BBA {} {Vourlidas}%
}{%
{\protect \APACyear {2009}}%
}]{%
ontiveros09}
\APACinsertmetastar {%
ontiveros09}%
\begin{APACrefauthors}%
{Ontiveros}, V.%
\BCBT {}\ \BBA {} {Vourlidas}, A.%
\end{APACrefauthors}%
\unskip\
\newblock
\APACrefYearMonthDay{2009}{Mar}{}.
\newblock
{\BBOQ}\APACrefatitle {{Quantitative Measurements of Coronal Mass
  Ejection-Driven Shocks from LASCO Observations}} {{Quantitative Measurements
  of Coronal Mass Ejection-Driven Shocks from LASCO Observations}}.{\BBCQ}
\newblock
\APACjournalVolNumPages{\apj}{693}{1}{267-275}.
\newblock
\begin{APACrefDOI} \doi{10.1088/0004-637X/693/1/267} \end{APACrefDOI}
\PrintBackRefs{\CurrentBib}

\bibitem [\protect \citeauthoryear {%
Owens%
}{%
Owens%
}{%
{\protect \APACyear {2018}}%
}]{%
owens18}
\APACinsertmetastar {%
owens18}%
\begin{APACrefauthors}%
Owens, M\BPBI J.%
\end{APACrefauthors}%
\unskip\
\newblock
\APACrefYearMonthDay{2018}{Aug}{23}.
\newblock
{\BBOQ}\APACrefatitle {Solar Wind and Heavy Ion Properties of Interplanetary
  Coronal Mass Ejections} {Solar wind and heavy ion properties of
  interplanetary coronal mass ejections}.{\BBCQ}
\newblock
\APACjournalVolNumPages{Solar Physics}{293}{8}{122}.
\newblock
\begin{APACrefURL} \url{https://doi.org/10.1007/s11207-018-1343-0}
  \end{APACrefURL}
\newblock
\begin{APACrefDOI} \doi{10.1007/s11207-018-1343-0} \end{APACrefDOI}
\PrintBackRefs{\CurrentBib}

\bibitem [\protect \citeauthoryear {%
{Pluta}%
, {Mrotzek}%
, {Vourlidas}%
, {Bothmer}%
\BCBL {}\ \BBA {} {Savani}%
}{%
{Pluta}%
\ \protect \BOthers {.}}{%
{\protect \APACyear {2019}}%
}]{%
pluta19}
\APACinsertmetastar {%
pluta19}%
\begin{APACrefauthors}%
{Pluta}, A.%
, {Mrotzek}, N.%
, {Vourlidas}, A.%
, {Bothmer}, V.%
\BCBL {}\ \BBA {} {Savani}, N.%
\end{APACrefauthors}%
\unskip\
\newblock
\APACrefYearMonthDay{2019}{Mar}{}.
\newblock
{\BBOQ}\APACrefatitle {{Combined geometrical modelling and white-light mass
  determination of coronal mass ejections}} {{Combined geometrical modelling
  and white-light mass determination of coronal mass ejections}}.{\BBCQ}
\newblock
\APACjournalVolNumPages{Astronomy and Astrophysics}{623}{}{A139}.
\newblock
\begin{APACrefDOI} \doi{10.1051/0004-6361/201833829} \end{APACrefDOI}
\PrintBackRefs{\CurrentBib}

\bibitem [\protect \citeauthoryear {%
{Pomoell}%
\ \BBA {} {Poedts}%
}{%
{Pomoell}%
\ \BBA {} {Poedts}%
}{%
{\protect \APACyear {2018}}%
}]{%
pomoell18}
\APACinsertmetastar {%
pomoell18}%
\begin{APACrefauthors}%
{Pomoell}, J.%
\BCBT {}\ \BBA {} {Poedts}, S.%
\end{APACrefauthors}%
\unskip\
\newblock
\APACrefYearMonthDay{2018}{Jun}{}.
\newblock
{\BBOQ}\APACrefatitle {{EUHFORIA: European heliospheric forecasting information
  asset}} {{EUHFORIA: European heliospheric forecasting information
  asset}}.{\BBCQ}
\newblock
\APACjournalVolNumPages{Journal of Space Weather and Space Climate}{8}{}{A35}.
\newblock
\begin{APACrefDOI} \doi{10.1051/swsc/2018020} \end{APACrefDOI}
\PrintBackRefs{\CurrentBib}

\bibitem [\protect \citeauthoryear {%
Richardson%
\ \BBA {} Cane%
}{%
Richardson%
\ \BBA {} Cane%
}{%
{\protect \APACyear {2010}}%
}]{%
richardson10}
\APACinsertmetastar {%
richardson10}%
\begin{APACrefauthors}%
Richardson, I\BPBI G.%
\BCBT {}\ \BBA {} Cane, H\BPBI V.%
\end{APACrefauthors}%
\unskip\
\newblock
\APACrefYearMonthDay{2010}{Jun}{01}.
\newblock
{\BBOQ}\APACrefatitle {Near-Earth Interplanetary Coronal Mass Ejections During
  Solar Cycle 23 (1996{\thinspace}--{\thinspace}2009): Catalog and Summary of
  Properties} {Near-earth interplanetary coronal mass ejections during solar
  cycle 23 (1996{\thinspace}--{\thinspace}2009): Catalog and summary of
  properties}.{\BBCQ}
\newblock
\APACjournalVolNumPages{Solar Physics}{264}{1}{189--237}.
\newblock
\begin{APACrefURL} \url{https://doi.org/10.1007/s11207-010-9568-6}
  \end{APACrefURL}
\newblock
\begin{APACrefDOI} \doi{10.1007/s11207-010-9568-6} \end{APACrefDOI}
\PrintBackRefs{\CurrentBib}

\bibitem [\protect \citeauthoryear {%
{Riley}%
\ \protect \BOthers {.}}{%
{Riley}%
\ \protect \BOthers {.}}{%
{\protect \APACyear {2018}}%
}]{%
riley18}
\APACinsertmetastar {%
riley18}%
\begin{APACrefauthors}%
{Riley}, P.%
, {Mays}, M\BPBI L.%
, {Andries}, J.%
, {Amerstorfer}, T.%
, {Biesecker}, D.%
, {Delouille}, V.%
\BDBL {}{Zhao}, X.%
\end{APACrefauthors}%
\unskip\
\newblock
\APACrefYearMonthDay{2018}{Sep}{}.
\newblock
{\BBOQ}\APACrefatitle {{Forecasting the Arrival Time of Coronal Mass Ejections:
  Analysis of the CCMC CME Scoreboard}} {{Forecasting the Arrival Time of
  Coronal Mass Ejections: Analysis of the CCMC CME Scoreboard}}.{\BBCQ}
\newblock
\APACjournalVolNumPages{Space Weather}{16}{9}{1245-1260}.
\newblock
\begin{APACrefDOI} \doi{10.1029/2018SW001962} \end{APACrefDOI}
\PrintBackRefs{\CurrentBib}

\bibitem [\protect \citeauthoryear {%
{Rouillard}%
}{%
{Rouillard}%
}{%
{\protect \APACyear {2011}}%
}]{%
rouillard11}
\APACinsertmetastar {%
rouillard11}%
\begin{APACrefauthors}%
{Rouillard}, A\BPBI P.%
\end{APACrefauthors}%
\unskip\
\newblock
\APACrefYearMonthDay{2011}{Jun}{}.
\newblock
{\BBOQ}\APACrefatitle {{Relating white light and in situ observations of
  coronal mass ejections: A review}} {{Relating white light and in situ
  observations of coronal mass ejections: A review}}.{\BBCQ}
\newblock
\APACjournalVolNumPages{Journal of Atmospheric and Solar-Terrestrial
  Physics}{73}{10}{1201-1213}.
\newblock
\begin{APACrefDOI} \doi{10.1016/j.jastp.2010.08.015} \end{APACrefDOI}
\PrintBackRefs{\CurrentBib}

\bibitem [\protect \citeauthoryear {%
{Ruffenach}%
\ \protect \BOthers {.}}{%
{Ruffenach}%
\ \protect \BOthers {.}}{%
{\protect \APACyear {2015}}%
}]{%
2015ruffenach}
\APACinsertmetastar {%
2015ruffenach}%
\begin{APACrefauthors}%
{Ruffenach}, A.%
, {Lavraud}, B.%
, {Farrugia}, C\BPBI J.%
, {D{\'e}moulin}, P.%
, {Dasso}, S.%
, {Owens}, M\BPBI J.%
\BDBL {}{Foullon}, C.%
\end{APACrefauthors}%
\unskip\
\newblock
\APACrefYearMonthDay{2015}{Jan}{}.
\newblock
{\BBOQ}\APACrefatitle {{Statistical study of magnetic cloud erosion by magnetic
  reconnection}} {{Statistical study of magnetic cloud erosion by magnetic
  reconnection}}.{\BBCQ}
\newblock
\APACjournalVolNumPages{Journal of Geophysical Research (Space
  Physics)}{120}{1}{43-60}.
\newblock
\begin{APACrefDOI} \doi{10.1002/2014JA020628} \end{APACrefDOI}
\PrintBackRefs{\CurrentBib}

\bibitem [\protect \citeauthoryear {%
{Sachdeva}%
}{%
{Sachdeva}%
}{%
{\protect \APACyear {2019}}%
}]{%
sachdeva19}
\APACinsertmetastar {%
sachdeva19}%
\begin{APACrefauthors}%
{Sachdeva}, N.%
\end{APACrefauthors}%
\unskip\
\newblock
\APACrefYearMonthDay{2019}{Jul}{}.
\newblock
{\BBOQ}\APACrefatitle {{Dynamics of solar Coronal Mass Ejections: forces that
  impact their propagation}} {{Dynamics of solar Coronal Mass Ejections: forces
  that impact their propagation}}.{\BBCQ}
\newblock
\APACjournalVolNumPages{arXiv e-prints}{}{}{arXiv:1907.12673}.
\PrintBackRefs{\CurrentBib}

\bibitem [\protect \citeauthoryear {%
{Sachdeva}%
, {Subramanian}%
, {Colaninno}%
\BCBL {}\ \BBA {} {Vourlidas}%
}{%
{Sachdeva}%
\ \protect \BOthers {.}}{%
{\protect \APACyear {2015}}%
}]{%
sachdeva15}
\APACinsertmetastar {%
sachdeva15}%
\begin{APACrefauthors}%
{Sachdeva}, N.%
, {Subramanian}, P.%
, {Colaninno}, R.%
\BCBL {}\ \BBA {} {Vourlidas}, A.%
\end{APACrefauthors}%
\unskip\
\newblock
\APACrefYearMonthDay{2015}{Aug}{}.
\newblock
{\BBOQ}\APACrefatitle {{CME Propagation: Where does Aerodynamic Drag 'Take
  Over'?}} {{CME Propagation: Where does Aerodynamic Drag 'Take Over'?}}{\BBCQ}
\newblock
\APACjournalVolNumPages{\apj}{809}{2}{158}.
\newblock
\begin{APACrefDOI} \doi{10.1088/0004-637X/809/2/158} \end{APACrefDOI}
\PrintBackRefs{\CurrentBib}

\bibitem [\protect \citeauthoryear {%
{Savani}%
\ \protect \BOthers {.}}{%
{Savani}%
\ \protect \BOthers {.}}{%
{\protect \APACyear {2013}}%
}]{%
savani13}
\APACinsertmetastar {%
savani13}%
\begin{APACrefauthors}%
{Savani}, N\BPBI P.%
, {Vourlidas}, A.%
, {Pulkkinen}, A.%
, {Nieves-Chinchilla}, T.%
, {Lavraud}, B.%
\BCBL {}\ \BBA {} {Owens}, M\BPBI J.%
\end{APACrefauthors}%
\unskip\
\newblock
\APACrefYearMonthDay{2013}{May}{}.
\newblock
{\BBOQ}\APACrefatitle {{Tracking the momentum flux of a CME and quantifying its
  influence on geomagnetically induced currents at Earth}} {{Tracking the
  momentum flux of a CME and quantifying its influence on geomagnetically
  induced currents at Earth}}.{\BBCQ}
\newblock
\APACjournalVolNumPages{Space Weather}{11}{5}{245-261}.
\newblock
\begin{APACrefDOI} \doi{10.1002/swe.20038} \end{APACrefDOI}
\PrintBackRefs{\CurrentBib}

\bibitem [\protect \citeauthoryear {%
{Siscoe}%
\ \BBA {} {Odstrcil}%
}{%
{Siscoe}%
\ \BBA {} {Odstrcil}%
}{%
{\protect \APACyear {2008}}%
}]{%
siscoe08}
\APACinsertmetastar {%
siscoe08}%
\begin{APACrefauthors}%
{Siscoe}, G.%
\BCBT {}\ \BBA {} {Odstrcil}, D.%
\end{APACrefauthors}%
\unskip\
\newblock
\APACrefYearMonthDay{2008}{Sep}{}.
\newblock
{\BBOQ}\APACrefatitle {{Ways in which ICME sheaths differ from magnetosheaths}}
  {{Ways in which ICME sheaths differ from magnetosheaths}}.{\BBCQ}
\newblock
\APACjournalVolNumPages{Journal of Geophysical Research (Space
  Physics)}{113}{A9}{A00B07}.
\newblock
\begin{APACrefDOI} \doi{10.1029/2008JA013142} \end{APACrefDOI}
\PrintBackRefs{\CurrentBib}

\bibitem [\protect \citeauthoryear {%
{Sudar}%
, {Vr{\v{s}}nak}%
\BCBL {}\ \BBA {} {Dumbovi{\'c}}%
}{%
{Sudar}%
\ \protect \BOthers {.}}{%
{\protect \APACyear {2016}}%
}]{%
sudar16}
\APACinsertmetastar {%
sudar16}%
\begin{APACrefauthors}%
{Sudar}, D.%
, {Vr{\v{s}}nak}, B.%
\BCBL {}\ \BBA {} {Dumbovi{\'c}}, M.%
\end{APACrefauthors}%
\unskip\
\newblock
\APACrefYearMonthDay{2016}{{\APACmonth{02}}}{}.
\newblock
{\BBOQ}\APACrefatitle {{Predicting coronal mass ejections transit times to
  Earth with neural network}} {{Predicting coronal mass ejections transit times
  to Earth with neural network}}.{\BBCQ}
\newblock
\APACjournalVolNumPages{\mnras}{456}{2}{1542-1548}.
\newblock
\begin{APACrefDOI} \doi{10.1093/mnras/stv2782} \end{APACrefDOI}
\PrintBackRefs{\CurrentBib}

\bibitem [\protect \citeauthoryear {%
{Takahashi}%
\ \BBA {} {Shibata}%
}{%
{Takahashi}%
\ \BBA {} {Shibata}%
}{%
{\protect \APACyear {2017}}%
}]{%
takahashi17}
\APACinsertmetastar {%
takahashi17}%
\begin{APACrefauthors}%
{Takahashi}, T.%
\BCBT {}\ \BBA {} {Shibata}, K.%
\end{APACrefauthors}%
\unskip\
\newblock
\APACrefYearMonthDay{2017}{Mar}{}.
\newblock
{\BBOQ}\APACrefatitle {{Sheath-accumulating Propagation of Interplanetary
  Coronal Mass Ejection}} {{Sheath-accumulating Propagation of Interplanetary
  Coronal Mass Ejection}}.{\BBCQ}
\newblock
\APACjournalVolNumPages{\apjl}{837}{2}{L17}.
\newblock
\begin{APACrefDOI} \doi{10.3847/2041-8213/aa624c} \end{APACrefDOI}
\PrintBackRefs{\CurrentBib}

\bibitem [\protect \citeauthoryear {%
{Tappin}%
}{%
{Tappin}%
}{%
{\protect \APACyear {2006}}%
}]{%
tappin06}
\APACinsertmetastar {%
tappin06}%
\begin{APACrefauthors}%
{Tappin}, S\BPBI J.%
\end{APACrefauthors}%
\unskip\
\newblock
\APACrefYearMonthDay{2006}{Feb}{}.
\newblock
{\BBOQ}\APACrefatitle {{The Deceleration of an Interplanetary Transient from
  the Sun to 5 Au}} {{The Deceleration of an Interplanetary Transient from the
  Sun to 5 Au}}.{\BBCQ}
\newblock
\APACjournalVolNumPages{\solphys}{233}{2}{233-248}.
\newblock
\begin{APACrefDOI} \doi{10.1007/s11207-006-2065-2} \end{APACrefDOI}
\PrintBackRefs{\CurrentBib}

\bibitem [\protect \citeauthoryear {%
{Temmer}%
, {Reiss}%
, {Nikolic}%
, {Hofmeister}%
\BCBL {}\ \BBA {} {Veronig}%
}{%
{Temmer}%
\ \protect \BOthers {.}}{%
{\protect \APACyear {2017}}%
}]{%
temmer17}
\APACinsertmetastar {%
temmer17}%
\begin{APACrefauthors}%
{Temmer}, M.%
, {Reiss}, M\BPBI A.%
, {Nikolic}, L.%
, {Hofmeister}, S\BPBI J.%
\BCBL {}\ \BBA {} {Veronig}, A\BPBI M.%
\end{APACrefauthors}%
\unskip\
\newblock
\APACrefYearMonthDay{2017}{Feb}{}.
\newblock
{\BBOQ}\APACrefatitle {{Preconditioning of Interplanetary Space Due to
  Transient CME Disturbances}} {{Preconditioning of Interplanetary Space Due to
  Transient CME Disturbances}}.{\BBCQ}
\newblock
\APACjournalVolNumPages{\apj}{835}{2}{141}.
\newblock
\begin{APACrefDOI} \doi{10.3847/1538-4357/835/2/141} \end{APACrefDOI}
\PrintBackRefs{\CurrentBib}

\bibitem [\protect \citeauthoryear {%
{Temmer}%
\ \protect \BOthers {.}}{%
{Temmer}%
\ \protect \BOthers {.}}{%
{\protect \APACyear {2011}}%
}]{%
temmer11}
\APACinsertmetastar {%
temmer11}%
\begin{APACrefauthors}%
{Temmer}, M.%
, {Rollett}, T.%
, {M{\"o}stl}, C.%
, {Veronig}, A\BPBI M.%
, {Vr{\v{s}}nak}, B.%
\BCBL {}\ \BBA {} {Odstr{\v{c}}il}, D.%
\end{APACrefauthors}%
\unskip\
\newblock
\APACrefYearMonthDay{2011}{{\APACmonth{12}}}{}.
\newblock
{\BBOQ}\APACrefatitle {{Influence of the Ambient Solar Wind Flow on the
  Propagation Behavior of Interplanetary Coronal Mass Ejections}} {{Influence
  of the Ambient Solar Wind Flow on the Propagation Behavior of Interplanetary
  Coronal Mass Ejections}}.{\BBCQ}
\newblock
\APACjournalVolNumPages{\apj}{743}{2}{101}.
\newblock
\begin{APACrefDOI} \doi{10.1088/0004-637X/743/2/101} \end{APACrefDOI}
\PrintBackRefs{\CurrentBib}

\bibitem [\protect \citeauthoryear {%
{Thernisien}%
}{%
{Thernisien}%
}{%
{\protect \APACyear {2011}}%
}]{%
thernisien11}
\APACinsertmetastar {%
thernisien11}%
\begin{APACrefauthors}%
{Thernisien}, A.%
\end{APACrefauthors}%
\unskip\
\newblock
\APACrefYearMonthDay{2011}{{\APACmonth{06}}}{}.
\newblock
{\BBOQ}\APACrefatitle {{Implementation of the Graduated Cylindrical Shell Model
  for the Three-dimensional Reconstruction of Coronal Mass Ejections}}
  {{Implementation of the Graduated Cylindrical Shell Model for the
  Three-dimensional Reconstruction of Coronal Mass Ejections}}.{\BBCQ}
\newblock
\APACjournalVolNumPages{\apjs}{194}{}{33}.
\newblock
\begin{APACrefDOI} \doi{10.1088/0067-0049/194/2/33} \end{APACrefDOI}
\PrintBackRefs{\CurrentBib}

\bibitem [\protect \citeauthoryear {%
{Thernisien}%
, {Vourlidas}%
\BCBL {}\ \BBA {} {Howard}%
}{%
{Thernisien}%
\ \protect \BOthers {.}}{%
{\protect \APACyear {2009}}%
}]{%
2009thernisien}
\APACinsertmetastar {%
2009thernisien}%
\begin{APACrefauthors}%
{Thernisien}, A.%
, {Vourlidas}, A.%
\BCBL {}\ \BBA {} {Howard}, R\BPBI A.%
\end{APACrefauthors}%
\unskip\
\newblock
\APACrefYearMonthDay{2009}{{\APACmonth{05}}}{}.
\newblock
{\BBOQ}\APACrefatitle {{Forward Modeling of Coronal Mass Ejections Using
  STEREO/SECCHI Data}} {{Forward Modeling of Coronal Mass Ejections Using
  STEREO/SECCHI Data}}.{\BBCQ}
\newblock
\APACjournalVolNumPages{\solphys}{256}{}{111-130}.
\newblock
\begin{APACrefDOI} \doi{10.1007/s11207-009-9346-5} \end{APACrefDOI}
\PrintBackRefs{\CurrentBib}

\bibitem [\protect \citeauthoryear {%
{Veronig}%
, {G{\"o}m{\"o}ry}%
, {Dissauer}%
, {Temmer}%
\BCBL {}\ \BBA {} {Vanninathan}%
}{%
{Veronig}%
\ \protect \BOthers {.}}{%
{\protect \APACyear {2019}}%
}]{%
veronig19}
\APACinsertmetastar {%
veronig19}%
\begin{APACrefauthors}%
{Veronig}, A\BPBI M.%
, {G{\"o}m{\"o}ry}, P.%
, {Dissauer}, K.%
, {Temmer}, M.%
\BCBL {}\ \BBA {} {Vanninathan}, K.%
\end{APACrefauthors}%
\unskip\
\newblock
\APACrefYearMonthDay{2019}{Jul}{}.
\newblock
{\BBOQ}\APACrefatitle {{Spectroscopy and Differential Emission Measure
  Diagnostics of a Coronal Dimming Associated with a Fast Halo CME}}
  {{Spectroscopy and Differential Emission Measure Diagnostics of a Coronal
  Dimming Associated with a Fast Halo CME}}.{\BBCQ}
\newblock
\APACjournalVolNumPages{\apj}{879}{2}{85}.
\newblock
\begin{APACrefDOI} \doi{10.3847/1538-4357/ab2712} \end{APACrefDOI}
\PrintBackRefs{\CurrentBib}

\bibitem [\protect \citeauthoryear {%
{Vr{\v s}nak}%
}{%
{Vr{\v s}nak}%
}{%
{\protect \APACyear {2001}}%
}]{%
vrsnak01}
\APACinsertmetastar {%
vrsnak01}%
\begin{APACrefauthors}%
{Vr{\v s}nak}, B.%
\end{APACrefauthors}%
\unskip\
\newblock
\APACrefYearMonthDay{2001}{{\APACmonth{11}}}{}.
\newblock
{\BBOQ}\APACrefatitle {{Dynamics of solar coronal eruptions}} {{Dynamics of
  solar coronal eruptions}}.{\BBCQ}
\newblock
\APACjournalVolNumPages{\jgr}{106}{}{25249-25260}.
\newblock
\begin{APACrefDOI} \doi{10.1029/2000JA004007} \end{APACrefDOI}
\PrintBackRefs{\CurrentBib}

\bibitem [\protect \citeauthoryear {%
{Vr{\v s}nak}%
, {Vrbanec}%
\BCBL {}\ \BBA {} {{\v C}alogovi{\'c}}%
}{%
{Vr{\v s}nak}%
\ \protect \BOthers {.}}{%
{\protect \APACyear {2008}}%
}]{%
vrsnak08a}
\APACinsertmetastar {%
vrsnak08a}%
\begin{APACrefauthors}%
{Vr{\v s}nak}, B.%
, {Vrbanec}, D.%
\BCBL {}\ \BBA {} {{\v C}alogovi{\'c}}, J.%
\end{APACrefauthors}%
\unskip\
\newblock
\APACrefYearMonthDay{2008}{{\APACmonth{11}}}{}.
\newblock
{\BBOQ}\APACrefatitle {{Dynamics of coronal mass ejections. The mass-scaling of
  the aerodynamic drag}} {{Dynamics of coronal mass ejections. The mass-scaling
  of the aerodynamic drag}}.{\BBCQ}
\newblock
\APACjournalVolNumPages{\aap}{490}{}{811-815}.
\newblock
\begin{APACrefDOI} \doi{10.1051/0004-6361:200810215} \end{APACrefDOI}
\PrintBackRefs{\CurrentBib}

\bibitem [\protect \citeauthoryear {%
{Vr{\v{s}}nak}%
}{%
{Vr{\v{s}}nak}%
}{%
{\protect \APACyear {2008}}%
}]{%
vrsnak08c}
\APACinsertmetastar {%
vrsnak08c}%
\begin{APACrefauthors}%
{Vr{\v{s}}nak}, B.%
\end{APACrefauthors}%
\unskip\
\newblock
\APACrefYearMonthDay{2008}{{\APACmonth{10}}}{}.
\newblock
{\BBOQ}\APACrefatitle {{Processes and mechanisms governing the initiation and
  propagation of CMEs}} {{Processes and mechanisms governing the initiation and
  propagation of CMEs}}.{\BBCQ}
\newblock
\APACjournalVolNumPages{Annales Geophysicae}{26}{10}{3089-3101}.
\newblock
\begin{APACrefDOI} \doi{10.5194/angeo-26-3089-2008} \end{APACrefDOI}
\PrintBackRefs{\CurrentBib}

\bibitem [\protect \citeauthoryear {%
{Vr{\v{s}}nak}%
}{%
{Vr{\v{s}}nak}%
}{%
{\protect \APACyear {2016}}%
}]{%
vrsnak16}
\APACinsertmetastar {%
vrsnak16}%
\begin{APACrefauthors}%
{Vr{\v{s}}nak}, B.%
\end{APACrefauthors}%
\unskip\
\newblock
\APACrefYearMonthDay{2016}{{\APACmonth{11}}}{}.
\newblock
{\BBOQ}\APACrefatitle {{Solar eruptions: The CME-flare relationship}} {{Solar
  eruptions: The CME-flare relationship}}.{\BBCQ}
\newblock
\APACjournalVolNumPages{Astronomische Nachrichten}{337}{10}{1002}.
\newblock
\begin{APACrefDOI} \doi{10.1002/asna.201612424} \end{APACrefDOI}
\PrintBackRefs{\CurrentBib}

\bibitem [\protect \citeauthoryear {%
{Vr{\v{s}}nak}%
\ \protect \BOthers {.}}{%
{Vr{\v{s}}nak}%
\ \protect \BOthers {.}}{%
{\protect \APACyear {2019}}%
}]{%
vrsnak19}
\APACinsertmetastar {%
vrsnak19}%
\begin{APACrefauthors}%
{Vr{\v{s}}nak}, B.%
, {Amerstorfer}, T.%
, {Dumbovi{\'c}}, M.%
, {Leitner}, M.%
, {Veronig}, A\BPBI M.%
, {Temmer}, M.%
\BDBL {}{Galvin}, A\BPBI B.%
\end{APACrefauthors}%
\unskip\
\newblock
\APACrefYearMonthDay{2019}{Jun}{}.
\newblock
{\BBOQ}\APACrefatitle {{Heliospheric Evolution of Magnetic Clouds}}
  {{Heliospheric Evolution of Magnetic Clouds}}.{\BBCQ}
\newblock
\APACjournalVolNumPages{\apj}{877}{2}{77}.
\newblock
\begin{APACrefDOI} \doi{10.3847/1538-4357/ab190a} \end{APACrefDOI}
\PrintBackRefs{\CurrentBib}

\bibitem [\protect \citeauthoryear {%
{Vr{\v{s}}nak}%
\ \protect \BOthers {.}}{%
{Vr{\v{s}}nak}%
\ \protect \BOthers {.}}{%
{\protect \APACyear {2010}}%
}]{%
vrsnak10}
\APACinsertmetastar {%
vrsnak10}%
\begin{APACrefauthors}%
{Vr{\v{s}}nak}, B.%
, {{\v{Z}}ic}, T.%
, {Falkenberg}, T\BPBI V.%
, {M{\"o}stl}, C.%
, {Vennerstrom}, S.%
\BCBL {}\ \BBA {} {Vrbanec}, D.%
\end{APACrefauthors}%
\unskip\
\newblock
\APACrefYearMonthDay{2010}{Mar}{}.
\newblock
{\BBOQ}\APACrefatitle {{The role of aerodynamic drag in propagation of
  interplanetary coronal mass ejections}} {{The role of aerodynamic drag in
  propagation of interplanetary coronal mass ejections}}.{\BBCQ}
\newblock
\APACjournalVolNumPages{\aap}{512}{}{A43}.
\newblock
\begin{APACrefDOI} \doi{10.1051/0004-6361/200913482} \end{APACrefDOI}
\PrintBackRefs{\CurrentBib}

\bibitem [\protect \citeauthoryear {%
{Vr{\v{s}}nak}%
\ \protect \BOthers {.}}{%
{Vr{\v{s}}nak}%
\ \protect \BOthers {.}}{%
{\protect \APACyear {2013}}%
}]{%
vrsnak13}
\APACinsertmetastar {%
vrsnak13}%
\begin{APACrefauthors}%
{Vr{\v{s}}nak}, B.%
, {{\v{Z}}ic}, T.%
, {Vrbanec}, D.%
, {Temmer}, M.%
, {Rollett}, T.%
, {M{\"o}stl}, C.%
\BDBL {}{Shanmugaraju}, A.%
\end{APACrefauthors}%
\unskip\
\newblock
\APACrefYearMonthDay{2013}{Jul}{}.
\newblock
{\BBOQ}\APACrefatitle {{Propagation of Interplanetary Coronal Mass Ejections:
  The Drag-Based Model}} {{Propagation of Interplanetary Coronal Mass
  Ejections: The Drag-Based Model}}.{\BBCQ}
\newblock
\APACjournalVolNumPages{\solphys}{285}{1-2}{295-315}.
\newblock
\begin{APACrefDOI} \doi{10.1007/s11207-012-0035-4} \end{APACrefDOI}
\PrintBackRefs{\CurrentBib}

\bibitem [\protect \citeauthoryear {%
{Wang}%
, {Shen}%
, {Wang}%
\BCBL {}\ \BBA {} {Ye}%
}{%
{Wang}%
\ \protect \BOthers {.}}{%
{\protect \APACyear {2004}}%
}]{%
wang04}
\APACinsertmetastar {%
wang04}%
\begin{APACrefauthors}%
{Wang}, Y.%
, {Shen}, C.%
, {Wang}, S.%
\BCBL {}\ \BBA {} {Ye}, P.%
\end{APACrefauthors}%
\unskip\
\newblock
\APACrefYearMonthDay{2004}{{\APACmonth{08}}}{}.
\newblock
{\BBOQ}\APACrefatitle {{Deflection of coronal mass ejection in the
  interplanetary medium}} {{Deflection of coronal mass ejection in the
  interplanetary medium}}.{\BBCQ}
\newblock
\APACjournalVolNumPages{\solphys}{222}{2}{329-343}.
\newblock
\begin{APACrefDOI} \doi{10.1023/B:SOLA.0000043576.21942.aa} \end{APACrefDOI}
\PrintBackRefs{\CurrentBib}

\bibitem [\protect \citeauthoryear {%
{Werner}%
, {Yordanova}%
, {Dimmock}%
\BCBL {}\ \BBA {} {Temmer}%
}{%
{Werner}%
\ \protect \BOthers {.}}{%
{\protect \APACyear {2019}}%
}]{%
werner19}
\APACinsertmetastar {%
werner19}%
\begin{APACrefauthors}%
{Werner}, A\BPBI L\BPBI E.%
, {Yordanova}, E.%
, {Dimmock}, A\BPBI P.%
\BCBL {}\ \BBA {} {Temmer}, M.%
\end{APACrefauthors}%
\unskip\
\newblock
\APACrefYearMonthDay{2019}{Feb}{}.
\newblock
{\BBOQ}\APACrefatitle {{Modeling the Multiple CME Interaction Event on 6-9
  September 2017 with WSA-ENLIL+Cone}} {{Modeling the Multiple CME Interaction
  Event on 6-9 September 2017 with WSA-ENLIL+Cone}}.{\BBCQ}
\newblock
\APACjournalVolNumPages{Space Weather}{17}{2}{357-369}.
\newblock
\begin{APACrefDOI} \doi{10.1029/2018SW001993} \end{APACrefDOI}
\PrintBackRefs{\CurrentBib}

\bibitem [\protect \citeauthoryear {%
{Wood}%
\ \protect \BOthers {.}}{%
{Wood}%
\ \protect \BOthers {.}}{%
{\protect \APACyear {2017}}%
}]{%
wood17}
\APACinsertmetastar {%
wood17}%
\begin{APACrefauthors}%
{Wood}, B\BPBI E.%
, {Wu}, C\BHBI C.%
, {Lepping}, R\BPBI P.%
, {Nieves-Chinchilla}, T.%
, {Howard}, R\BPBI A.%
, {Linton}, M\BPBI G.%
\BCBL {}\ \BBA {} {Socker}, D\BPBI G.%
\end{APACrefauthors}%
\unskip\
\newblock
\APACrefYearMonthDay{2017}{Apr}{}.
\newblock
{\BBOQ}\APACrefatitle {{A STEREO Survey of Magnetic Cloud Coronal Mass
  Ejections Observed at Earth in 2008-2012}} {{A STEREO Survey of Magnetic
  Cloud Coronal Mass Ejections Observed at Earth in 2008-2012}}.{\BBCQ}
\newblock
\APACjournalVolNumPages{\apjs}{229}{2}{29}.
\newblock
\begin{APACrefDOI} \doi{10.3847/1538-4365/229/2/29} \end{APACrefDOI}
\PrintBackRefs{\CurrentBib}

\bibitem [\protect \citeauthoryear {%
{Yashiro}%
\ \protect \BOthers {.}}{%
{Yashiro}%
\ \protect \BOthers {.}}{%
{\protect \APACyear {2004}}%
}]{%
yashiro04}
\APACinsertmetastar {%
yashiro04}%
\begin{APACrefauthors}%
{Yashiro}, S.%
, {Gopalswamy}, N.%
, {Michalek}, G.%
, {St. Cyr}, O\BPBI C.%
, {Plunkett}, S\BPBI P.%
, {Rich}, N\BPBI B.%
\BCBL {}\ \BBA {} {Howard}, R\BPBI A.%
\end{APACrefauthors}%
\unskip\
\newblock
\APACrefYearMonthDay{2004}{Jul}{}.
\newblock
{\BBOQ}\APACrefatitle {{A catalog of white light coronal mass ejections
  observed by the SOHO spacecraft}} {{A catalog of white light coronal mass
  ejections observed by the SOHO spacecraft}}.{\BBCQ}
\newblock
\APACjournalVolNumPages{Journal of Geophysical Research (Space
  Physics)}{109}{A7}{A07105}.
\newblock
\begin{APACrefDOI} \doi{10.1029/2003JA010282} \end{APACrefDOI}
\PrintBackRefs{\CurrentBib}

\end{thebibliography}
 
\end{document}